\documentclass[showpacs,preprintnumbers]{revtex4}
\pdfoutput=1
\usepackage{amsfonts}
\usepackage{bm}
\usepackage{graphicx}
\usepackage{amsmath}
\usepackage{amssymb}
\usepackage{times}
\usepackage{braket}
\let\mathbf=\boldsymbol

\begin{document}
\title{Domain wall of a ferromagnet on a three-dimensional topological insulator}
\author{Ryohei Wakatsuki$^{1}$}
\email{wakatsuki@appi.t.u-tokyo.ac.jp}
\author{Motohiko Ezawa$^{1}$}
\author{Naoto Nagaosa$^{1,2}$}
\affiliation{$^{1}$ Department of Applied Physics, University of Tokyo, 7-3-1, Hongo, Bunkyo-ku, Tokyo 113-8656, Japan}
\affiliation{$^{2}$ RIKEN Center for Emergent Matter Science (CEMS), Wako, Saitama 351-0198, Japan}

\begin{abstract}
\textbf{Topological insulators (TIs) show rich phenomena and functions which can never be realized in ordinary insulators. Most of them come from the peculiar surface or edge states. Especially, the quantized anomalous Hall effect (QAHE) without an external magnetic field is realized in the two-dimensional ferromagnet on a three-dimensional TI which supports the dissipationless edge current. Here we demonstrate theoretically that the domain wall of this ferromagnet, which carries edge current, is charged and can be controlled by
the external electric field. The chirality and relative stability of the Neel wall and Bloch wall depend on the position of the Fermi energy as well as the form of the coupling between the magnetic moments and orbital of the host TI. These findings will pave a path to utilize the magnets on TI for the spintronics applications. }

\end{abstract}
\maketitle

The dissipationless topological currents (TIs) are the issue of current
great interests. TIs  and superconductors are the two
representative materials which support the dissipationless currents on their
surface~\cite{TI1,TI2}. 
These materials are characterized by the gapped bulk states and
gapless surface or edge states due to bulk--edge or bulk--surface
correspondence. The surface Weyl states of a three-dimensional (3D) TI
offer an arena for various novel physical properties
due to its momentum--spin locking, as described by the two-dimensional (2D)
Hamiltonian,
\begin{equation}
H=\pm v_{\text{F}}(\mathbf{e}_{z}\times\mathbf{p})\cdot\mathbf{\sigma},
\end{equation}
where $\mathbf{e}_{z}$ is the normal unit vector to the surface,
$\mathbf{\sigma}=(\sigma_{x},\sigma_{y},\sigma_{z})$ are the Pauli matrices,
and $\mathbf{p}$ is the 2D momentum. The sign $\pm$ differs for the top and
bottom surfaces. 

This surface state shows various unique properties when magnetic moments 
are coupled to it. 
For example, the effect of the doped magnetic moments on the transport 
properties has been studied theoretically~\cite{FCZhang}.
Another remarkable phenomena is the quantized anomalous Hall effect (QAHE), 
where the Hall conductance $\sigma_{xy}$ is quantized with the vanishing 
longitudinal conductance without
the external magnetic field~\cite{Haldane,Onoda,Yu,Nomura,Abanin}. 
When the exchange coupling to the magnetization
is introduced, the Hamiltonian reads
\begin{equation}
H=\pm v_{\text{F}}(\mathbf{e}_{z}\times\mathbf{p})\cdot\mathbf{\sigma
}+J\mathbf{n}\cdot\mathbf{\sigma},\label{eq:Weyl}
\end{equation}
where $J$ is the exchange energy, and $\mathbf{n}$ is the direction of the
magnetization. When the magnetization is normal to the surface, i.e.,
$\mathbf{n}\parallel\mathbf{e}_{z}$, the the mass gap opens in the surface
state and half-quantized Hall conductance 
$\sigma_{xy}=\pm\frac{e^{2}}{2h}$, 
i.e., the quantized anomalous Hall effect (QAHE) is realized,
when the Fermi energy is tuned within in this mass gap. Note that the observed
Hall conductance is the sum of the upper and bottom surfaces and hence $\pm\frac{e^{2}}{h}$. 

The dynamics of the magnetization on 3D TI has been also studied theoretically based on the 2D Weyl Hamiltonian~\cite{Franz,Yokoyama,Loss,Linder,Ferreiros,Baum,Mendler,Hurst}.
Experimentally, the gap opening in the surface states of a 3D TI Bi$_2$Se$_3$ due to the doping of magnetic ions has been observed by angle-resolved photoemission spectroscopy (ARPES) ~\cite{Chen}. Also the QAHE has been recently 
observed in Bi$_{2}$Te$_{3}$
with Cr doping~\cite{Zhang,Chang,Joe,Chang2,Bestwick,Kou,Figuerou,Ni}. When the magnetization is along the
$z$ direction both for the top and bottom surfaces, the edge channel goes
along the side surface. The edge channel appears also along
the domain wall which separates the two domains of $\sigma_{xy}=\frac{e^{2}}{2h}$ 
and $\sigma_{xy}=-\frac{e^{2}}{2h}$. 

In the field of spintronics, the magnetic domain walls play important roles 
as the information carriers and their manipulation is a keen issue. 
Especially, the racetrack memory using the current-driven motion of the domain wall 
is proposed ~\cite{Parkin1}.  Recently, the vital role of the spin--orbit interaction (SOI)
in the domain wall motion has been revealed~\cite{Parkin2}. 
The spin-to-charge conversion by the SOI is 
also a hot topic in spintronics~\cite{Fert}.
Therefore, it is an important issue to
examine theoretically the domain walls in the ferromagnet on a TI
from the viewpoint of the spintronics, since the 
momentum--spin locking at the surface state
of the TI corresponds to the strong-coupling limit of the SOI.

There are some subtle issues in the Hamiltonian Eq. (\ref{eq:Weyl}): (i) One
needs to introduce the energy cut off to avoid the ultra-violet divergence,
which is naturally given by the band gap of the 3D bulk states; namely, the
surface states merge into the bulk conduction and valence bands. However, when
the in-plane components of the magnetization $n_{x},n_{y}$ are finite, the 2D
momentum $\mathbf{p}$ shifts, and the surface states near the merging points
are changed, which contribute to the energy but can not be properly described
by Eq. (\ref{eq:Weyl}). (ii) The exchange coupling to the magnetization in
Eq. (\ref{eq:Weyl}) needs to be re-examined.
The Cr atoms replaces Bi atoms, and can have the exchange coupling to the $p$-orbitals of both Bi and Te, but
with different weight. This changes the effective Hamiltonian for the surface state. (iii) The dependence on the depth of the magnetic layer, and the
relation between the top and bottom surfaces are of interest as well, which is
accessed only by the 3D model with finite thickness.

\begin{figure}[t]
\centerline{\includegraphics[width=0.49\textwidth]{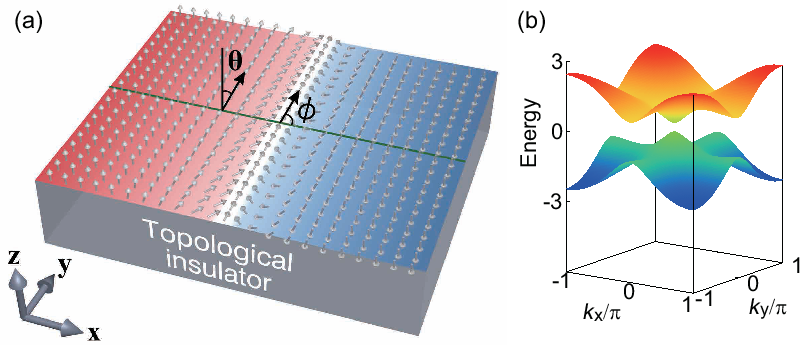}}\caption{(a) Illustration of the domain wall in the ferromagnet on a TI. Along the domain wall, the gapless chiral edge channel appears (white stripe region). The angle $\phi$ specifies the type of the domain wall, i.e., $\phi=0,\pi$ corresponds to Neel wall while $\phi=\pi/2,3 \pi/2$ to Bloch wall. (b) The surface band structure with homogeneous ferromagnetic calculated from the 3D tight-binding model. The vertical axis is the energy in unit of $t$.}
\label{FigWall}
\end{figure}

In this paper, we investigate the stability and charging effects of a domain
wall on the surface of the 3D TI based on the 3D
tight-binding model. We carry out a numerical study based on the 3D
tight-binding model~\cite{Liu,Shan,ZhangH}. We also perform an analytical study
based on the effective 2D surface Hamiltonian which we derive from the 3D
model. The exchange coupling is found to be anisotropic due to the orbital
dependence, as we have mentioned. Figure \ref{FigWall} shows the schematic structure of the domain wall on a TI. The angle $\phi$ determines the structure of the domain wall, i.e., Neel or Bloch wall and its chirality. It is found that the most stable domain wall structure depends on the position of the Fermi energy, i.e., one can control the domain structure by gating.
Another important result is that the domain wall is charged
due to the two effects: One originates in the zero-energy edge state along the
domain wall and the other in the charging effect of the magnetic texture. It
will offer a way to manipulate the domain wall by electric field.

\section*{Results}

\textbf{Model Hamiltonian.}
We start with the following minimum model for 3D TIs ~\cite{Liu,ZhangH},
\begin{equation}
H_{\text{3D}}=\hbar v_\text{F} \mathbf{k}\cdot\mathbf{\sigma}\tau_x+\left(m_0+m_2k^2\right)\tau_z + \left(J_0+J_3\tau_z\right) \mathbf{n}\cdot\mathbf{\sigma}, \label{continuum}
\end{equation}
where $v_F$ is the Fermi velocity, $m_0$ and $m_2$ are the mass parameters.
For the numerical calculation, we use the corresponding lattice model ~\cite{Liu,Shan,ZhangH},
\begin{equation}
H_{\text{3D}}=t\sum_{\alpha=x,y,z}\sin k_{\alpha}\sigma_{\alpha}\tau_{x}+m\left(  k\right)  \tau
_{z}+(J_0+J_3\tau_{z})\mathbf{n}\cdot\mathbf{\sigma},\label{3DHamil}
\end{equation}
where $t$ is the transfer integral, $\mathbf{\sigma}=(\sigma_{x},\sigma_{y},\sigma_{z})$ and $\mathbf{\tau}=(\tau_{x},\tau_{y},\tau_{z})$ are the Pauli matrices for the spin and pseudospin degrees of freedom, and
\begin{equation}
m\left(  k\right)  \equiv m_{0}+ 2 m_2 \left(  3-\cos k_{x}-\cos k_{y}-\cos k_{z}\right)  .
\end{equation}
The pseudospin represents the $p$-orbitals of the Bi and Te. We have
introduced an orbital-dependent exchange interaction, i.e., the $\tau_{z}=1$
orbital is coupled with $(J_0+J_3)\mathbf{n}\cdot\mathbf{\sigma}$, while the
$\tau_{z}=-1$ orbital with $(J_0-J_3)\mathbf{n}\cdot\mathbf{\sigma}$. When $J_3=\pm
J_0$, the exchange interactions exist only at one orbital, while equally coupled
when $J_3=0$.
In the case of Cr doped (Bi,Sb)$_{2}$Te$_{3}$, the magnetization is induced by
the substitution of the (Bi,Sb) atoms by the Cr atoms, which is coupled mostly to
the Te atoms. Hence it is expected that $J_3 \sim J_0$~\cite{Henk}.
Therefore, we consider the two limiting cases of $J_3=0$ and $J_3=J_0$.
The case $J_3=0$ is useful since it provides us with a clear physical picture from the analytical point of view.

The system without the magnetism is known~\cite{Liu,Shan,ZhangH} to be a strong
TI for $-12<m_{0}/m_{2}<-8$ and $-4<m_{0}/m_{2}<0$, a weak
TI for $-8<m_{0}/m_{2}<-4$, and the trivial insulator for
$m_{0}/m_{2}<-12$ and $0<m_{0}/m_{2}$. The strong TI phase
is the most intriguing, and hence we choose $m_{0}=-0.8$, $m_{2}=0.4$ and
$t=1$ for numerical calculations and for illustration throughout the paper.
The bulk gap is given by $2m_{0}$. Note that even the $4\times4$ tight-binding
Hamiltonian Eq. (\ref{3DHamil}) is an effective one around the top of the
valence band and the bottom of the conduction band. In actual materials, there
are many other bands which contribute to the higher energy and short
wavelength physics. Therefore, we regard the \textquotedblleft lattice
constant\textquotedblright\ (which is put to be unity) as the coarse grained one.

We are interested in the low-energy physics on the surface of the above TI.
We consider a slab geometry with finite thickness along the $z$ direction.
Then, the 2D Weyl fermions appear both on the top and bottom surfaces.
This can be seen from the 2D low-energy Hamiltonian by projecting the 3D continuum Hamiltonian (\ref{continuum}) onto the space spanned by the surface states.
The result is
\begin{equation}
H_{\text{2D}}=\pm \hbar v_\text{F} \left(\mathbf{e}_z \times \mathbf{k}\right) \cdot \mathbf{\sigma} + J_\parallel \left(n_x\sigma_x+n_y \sigma_y\right)+J_\perp n_z\sigma_z, \label{2DHamil}
\end{equation}
with the parameters $J_\parallel$ and $J_\perp$, which are related with $J_0$ and $J_3$ in Eq. (\ref{continuum}) and (\ref{3DHamil}).
It can be derived as follows.

At the $\Gamma$ point, we obtain the surface states by solving the eigenequation (\ref{continuum}) without the exchange terms by setting $k_x=k_y=0$ and $k_z\rightarrow -i\partial_z$ for the semi-infinite system.
The top and bottom surface states are represented as
\begin{equation}
\ket{\psi_s}=\phi\left(z\right)\ket{\sigma_z=s}\otimes\ket{\tau_y=\pm s},
\end{equation}
where $\pm$ in the $\tau$ part corresponds to the top and bottom surface respectively, and $s=\pm1$ represents the spin eigenvalue.
Therefore, we find
\begin{equation}
\bra{\psi_s}\sigma_{x}\ket{\psi_{s'}}=0,\quad
\bra{\psi_s}\sigma_{y}\ket{\psi_{s'}}=0,\quad
\bra{\psi_s}\sigma_{z}\ket{\psi_{s'}}=\left(  \sigma_{z}\right)  _{s,s^{\prime}}.
\end{equation}
It follows that $J_{\perp}=J_{0}$, namely the exchange term in the 3D bulk Hamiltonian is projected into the Ising interaction in the 2D surface Hamiltonian.
For the orbital-dependent exchange term, the components are
\begin{equation}
\bra{\psi_s}\sigma_{x}\tau_{z}\ket{\psi_{s'}}=\left( \sigma_{x}\right)_{s,s^{\prime}},\quad
\bra{\psi_s}\sigma_{y}\tau_{z}\ket{\psi_{s'}}=\left( \sigma_{y}\right)_{s,s^{\prime}},\quad
\bra{\psi_s}\sigma_{z}\tau_{z}\ket{\psi_{s'}}=0.
\end{equation}
It follows that $J_{\parallel}=J_{3}$, namely the orbital-dependent exchange term in the 3D bulk Hamiltonian is projected into the in-plane exchange term in the 2D surface Hamiltonian.

Our important observation is that the exchange interaction on the 2D surface is anisotropic even if that in the 3D bulk is isotropic.
The perpendicular exchange interaction is induced by the $J_0$-term while the in-plane exchange interaction is induced by the $J_3$-term.

In what follows we carry out an analysis of the surface states of the
TI numerically based on the 3D Hamiltonian Eq. (\ref{3DHamil})
and analytically based on the 2D Hamiltonian Eq. (\ref{2DHamil}). The momentum
$k_{y}$ is a good quantum number since the surface are assumed to be
uniform in the $y$ direction. We numerically diagonalize the system with $128$ sites along the $x$ direction and $8$ sites along the $z$ direction for each $k_{y}$. We take $200$ points for $k_{y}$.
We set two domain walls to apply the periodic boundary condition for $x$ direction, and we illustrate figures for one of the domain walls throughout the paper.

\textbf{Magnetic domain wall.}
We consider a magnetic domain wall between the two degenerate ground states, $\bm{n}=\pm(0,0,1)$ lying along the $y$ axis on the surface of the TI,
\begin{equation}
\mathbf{n}\left(  x\right)  =\left(  \sin\theta (x) \cos\phi,\sin\theta (x) \sin\phi ,\cos\theta (x) \right).\label{DW}
\end{equation}
with $\cos\theta (x)=\tanh\frac{x}{\xi}$. 
The angle $\phi$ represents the type of magnetic domain wall.
Especially, $\phi=0,\pi$ represent the Neel walls, while $\phi=\pi/2,3\pi/2$ the Bloch walls,
\begin{equation}
\mathbf{n}_{\text{N}}=\left(  \pm\text{sech}\frac{x}{\xi},0,\tanh\frac{x}{\xi
}\right)  ,\quad\mathbf{n}_{\text{B}}=\left(  0,\pm\text{sech}\frac{x}{\xi
},\tanh\frac{x}{\xi}\right)  .\label{DWNB}
\end{equation}
We call $\phi=0$ ($\phi=\pi$) as Neel 1 (Neel 2), and $\phi=\pi/2, 3\pi/2$ as Bloch.
These two types of Bloch walls are related by the mirror symmetry operation with respect to the $zx$ plane. 

The domain wall width $\xi$ should be optimized as a variational parameter in Eq. (\ref{DW}). It
is found that the energy is decreased as $\xi$ is decreased down to $\xi=2.0$. See Supplementary Information \ref{WDep}.
Therefore, the width of the domain wall is typically the lattice constant in this
model. The reason is basically that the kinetic energy in the 2D effective
Hamiltonian is solely given by the SOI and hence
there is no length scale due to the SOI other than the lattice constant.
The detailed discussion is given in Supplementary Information \ref{EffecAct}.
However, as mentioned above, the lattice constant of the present tight-binding Hamiltonian is that of the coarse grained model,
and hence the distinction between Neel and Bloch walls still makes sense. Also the
width depends on the additional single-ion magnetic anisotropy term
$Kn_{z}^{2}$ which exists in the real material but not included in the present model.
We have numerically confirmed that the qualitative features of the results do
not depend on $\xi$, and hence we have shown the results for $\xi=4.0$ for
illustrative purpose in order to clearly show the difference between the Neel and
Bloch walls.
See \ref{WDep} in Supplementary Information.

\begin{figure}[t]
\centerline{\includegraphics[width=0.48\textwidth]{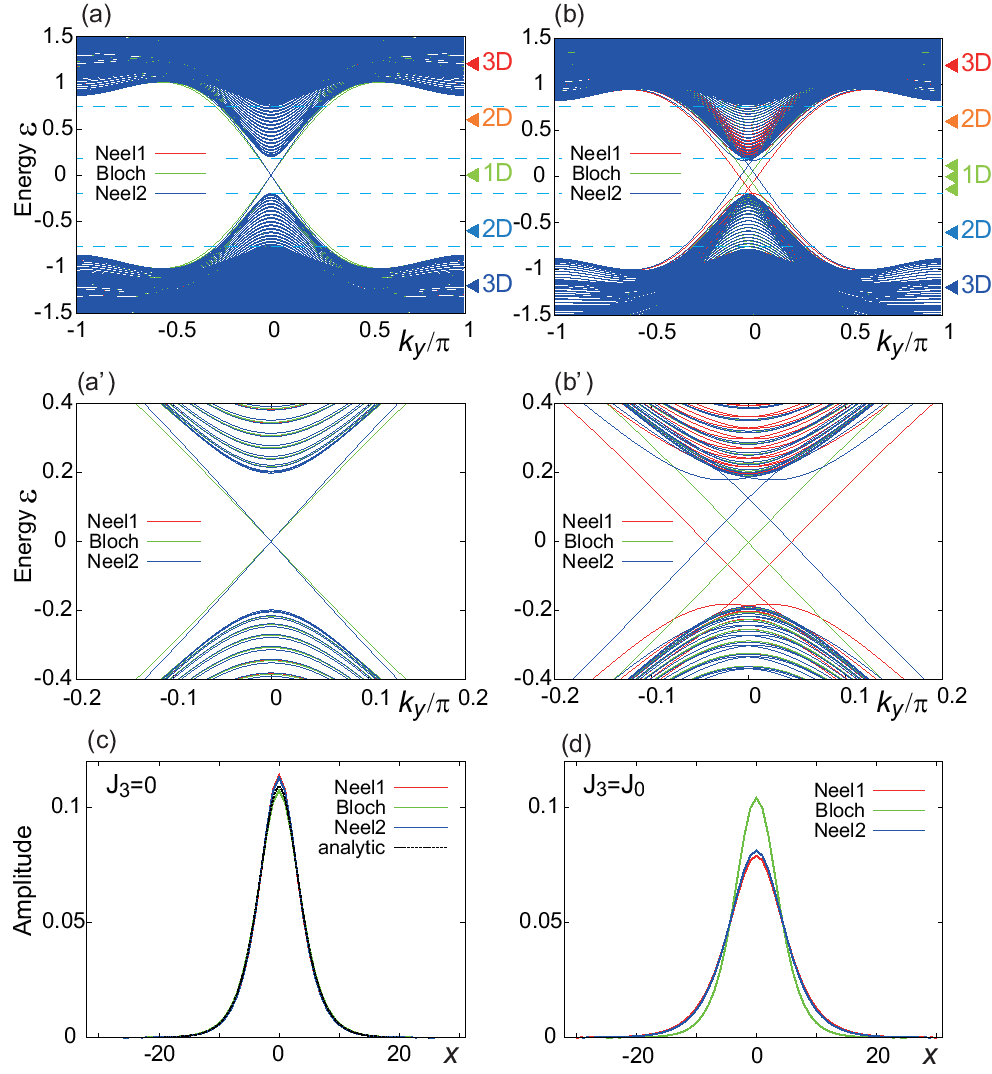}}\caption{The energy dispersions of the edge modes
for (a),(a') $J_3=0$ and (b),(b') $J_3=J_0$. The edge mode appears inside the bulk band
gap. The dispersion is almost independent of $\phi$, i.e., the type of domain wall, for $J_3=0$ ((a),(a')), while it is sensitive when $J_3=J_0$ ((b),(b')). Probability distribution of the zero-energy wave function for $k_y=0$ for (c) $J_3=0$, (d) $J_3=J_0$.
The dotted curve in (c) is given by Jackiw--Rebbi solution Eq. (\ref{ModeJR}) in the text.}
\label{FigEdge}
\end{figure}

\textbf{Edge modes.} A magnetic domain wall separates the two domains with up
and down spins, i.e., the regions of $\sigma_{xy}=\pm\frac{e^{2}}{2h}$.
Therefore, the difference of $\sigma_{xy}$ is $\frac{e^{2}}{h}$ and hence one chiral edge channel is expected to appear along the domain wall.
We show the energy dispersion and the probability distribution of the edge channel wave function along the $x$ direction obtained numerically for $J_3=0 $ in Figs. \ref{FigEdge}(a) and (a'), and $J_3=J_0$ in Figs.\ref{FigEdge}(b) and (b'), respectively.

There are three energy scales in the band structure as shown in
Figs. \ref{FigEdge}(a) and (b). One is the 3D bulk band structure which exists for
$|\varepsilon|>\left\vert m_{0}\right\vert $. The second is the 2D surface band
structure which exists for $|J_{\perp}|<|\varepsilon|<\left\vert m_{0}\right\vert $.
The last is the 1D edge states along the domain wall which exists for
$|\varepsilon|<|J_{\perp}|$.

When $J_3=0$, the dispersion and wave function of  the edge modes are almost independent of $\phi$ as shown in Figs. \ref{FigEdge}(a) and (a').
This is consistent with Eq. (\ref{2DHamil}) with $J_{\parallel}\propto J_3=0$.
Since the coupling is Ising-like, there is no $\phi$ dependence for the surface states.
We have determined numerically the probability distribution of the wave function at $k_{y}=0$, which we show in Figs. \ref{FigEdge}(c) and (d).

We present a clear physical picture for the zero-energy edge mode for $J_\parallel=J_3=0$.
The wave function is analytically given by the Jackiw--Rebbi solution~\cite{JR},
\begin{equation}
\psi_{\text{JR}}^{\uparrow}\left(  x\right)  = \psi_{\text{JR}}^{\downarrow
}\left(  x\right) = C\left( \cosh \frac{x}{\xi} \right)^{-J_\perp\frac{ \xi}{\hbar v_\text{F}}}, \label{ModeJR}
\end{equation}
with a normalization constant $C$.
It can be obtained by solving the differential equation given from Eq. (\ref{2DHamil})
\begin{equation}
\left[ -i \hbar v_\text{F} \sigma_y \partial_x + J_\perp n_z \left(x\right) \sigma_z \right] \psi \left(x\right) = 0.
\end{equation}
Indeed, it well explains the numerical data in Fig. \ref{FigEdge}(c).
The half width of the wave function is the same order of the domain wall width $\xi$.

On the other hand, the edge modes depend on $\phi$ when $J_3=J_0$ as shown in Figs. \ref{FigEdge}(b) and (b'). This is again consistent with Eq. (\ref{2DHamil}) with $J_{\parallel}=J_{\perp}$.
The energy dispersion of the edge mode is well described by
\begin{equation}
E\left(  k_{y}\right)  =\pm \hbar v_{\text{F}}k_{y}-\frac{J_{\parallel} \hbar v_{\text{F}}\Gamma\left(  \frac{1}{2}+\frac{J_{\perp}\xi}{\hbar v_{\text{F}}}\right)  ^{2}}{J_{\perp}\xi\Gamma\left(\frac{J_{\perp}\xi}{\hbar v_{\text{F}}}\right)  ^{2}} \cos \phi, \label{edge}
\end{equation}
as we derive by the first-order perturbation in Supplementary Information \ref{EdgeChan}.
Note that the spatial extent of the wave function is affected by the energy separation between the in-gap state and the edge of the bulk density states, and hence depends on $\phi$ in this case. 

\begin{figure}[t]
\centerline{\includegraphics[width=0.5\textwidth]{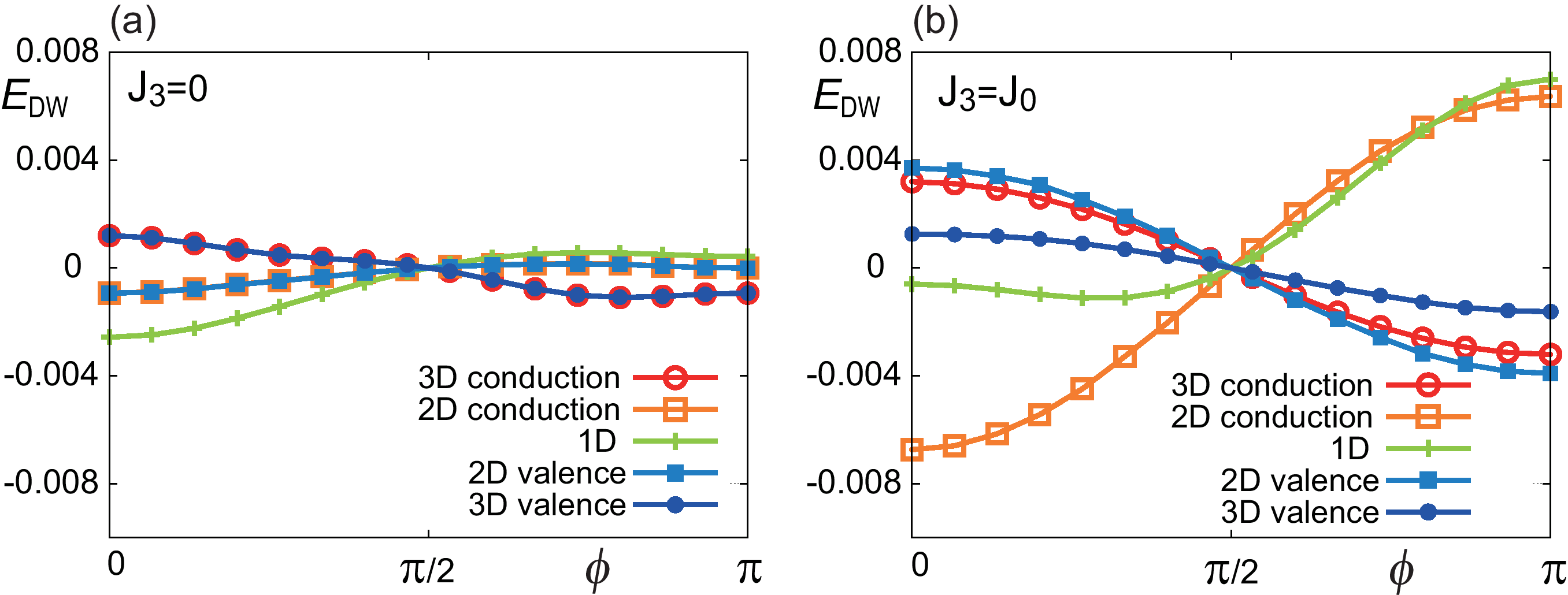}}\caption{The energy of the domain wall $E_{\text{DW}}$ as a function of $\phi$ for (a) $J_3=0$, (b) $J_3=J_0$. 
When $J_3=0$, the lowest energy domain wall structure is at $\phi=0$ for $\mu$  in the 2D valence/conduction bands or inside the gap. It turns into $\phi=\pi$ when $\mu$ is in the 3D valence/conduction bands. When $J_3=J_0$, on the other hand, $\phi=0$ is most stable when $\mu$ is in the 2D conduction band, and nearly Bloch wall $\phi \cong 0$ is stable for $\mu$ inside the gap. $\phi=\pi$ is the most stable for other cases. This behavior can be understood by considering the DM derived from the 2D surface states.}
\label{FigEnergy}
\end{figure}

\textbf{Domain wall energy.} We show in Fig. \ref{FigEnergy} the $\phi
$-dependence of the domain wall energy $E_{\text{DW}}$ measured from the value
at $\phi=\pi/2$ (Bloch wall) for several values of the chemical potential
$\mu$ when the magnetic layer is at the top surface. (The absolute value of
the domain wall energy compared with the uniform magnetization is a more
subtle quantity, which depends also on the magnetic anisotropy term
$Kn_{z}^{2}$, and therefore we do not address it in this paper.)
The domain wall energy is the same for $\phi$ and $2\pi-\phi$ due to the mirror symmetry with respect to $zx$ plane as $\sigma_{y}\mapsto-\sigma_{y}$, $n_{y}\mapsto-n_{y}$.
Therefore, it is enough to show the results for $0\leq\phi\leq\pi$.
$E_{\text{DW}}(\phi)$ behaves quite differently between the cases of $J_3=0$ and $J_3=J_0$.
(In Supplementary Information \ref{alphadep}, Fig. \ref{alpha} illustrates $E_\text{DW}$ for various values of $J_3$.)
When $J_3=0$, the Neel wall with $\phi=0$ is the most stable for the chemical potential $\mu$  in the 2D valence/conduction bands or inside the gap. When $\mu$ is in the 3D bands, the Neel wall with $\phi=\pi$ becomes the most stable.
In this case, the system possesses the particle--hole symmetry as shown in Supplementary Information \ref{SymmAna}.
As a result, the energy is symmetric between $\mu\longleftrightarrow-\mu$, which is also verified by our numerical calculations.

\begin{figure}[t]
\centerline{\includegraphics[width=0.5\textwidth]{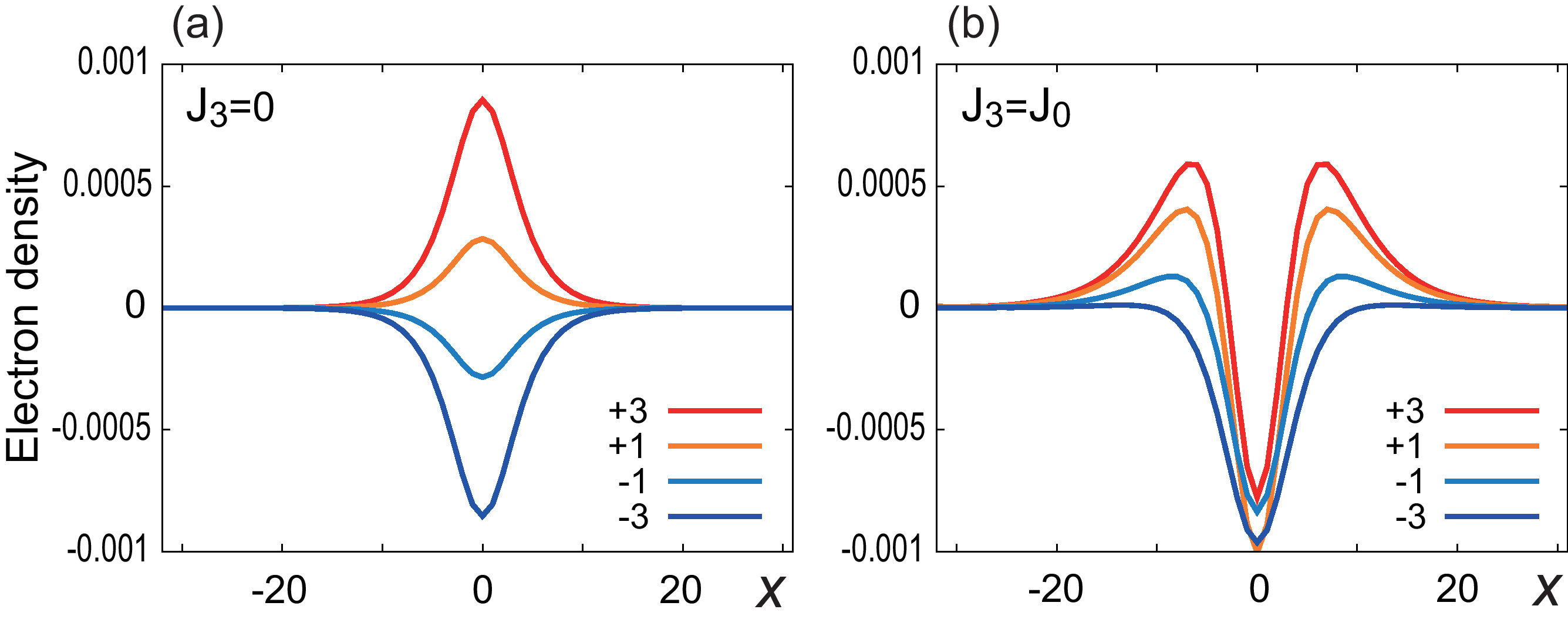}}\caption{Electron density distribution of the optimized domain wall structure for various chemical potential for (a) $J_3=0$ and (b) $J_3=J_0$.
Electrons (holes) are localized at the zero-energy states due to the magnetic
domain wall for $\mu>0$ ($\mu<0$). The numbers $\pm 1$, $\pm 3$ indicate the electron number measured from the half-filling. The horizontal axis is the $x$ coordinate.}
\label{FigCharge}
\end{figure}

On the other hand, when $J_3\neq0$, the particle--hole symmetry is lost.
For $J_3=J_0$ in Fig. \ref{FigEnergy}(b), the minimum energy configuration changes
from $\phi=\pi$ (Neel 2) for large positive $\mu>0$ (in the 3D conduction band), turns to $\phi=0$ (Neel1) for $\mu$ in the 2D conduction band, approaches to $\phi=\pi/2$ (Bloch) for $\mu$ within the gap, and eventually to $\phi=\pi$ (Neel 2) for $\mu<0$.
This means that one can control the angle $\phi$ of the domain wall by the gate voltage, which changes
the chemical potential. This is one of our main results in the present paper.
This change of the stable magnetic structure is understood analytically in
terms of the effective Dzyaloshinskii--Moriya (DM) interaction induced from
the TI surface state as discussed below.

When the chemical potential is in the 2D surface band $|J_\perp |<|\mu|<|m_{0}|$, the stability of a magnetic domain wall can be understood in terms of the effective surface DM interaction due to the Weyl surface states.
In order to derive the effective Hamiltonian for the magnets, we integrate out the fermion degrees of freedom, namely, calculate the following effective action
\begin{equation}
S_{\text{eff}}=\frac{1}{2} \sum_{\alpha,\beta=x,y,z} J_\alpha J_\beta \sum_{\mathbf{q}}n^{\alpha}(\mathbf{q})\chi^{\alpha \beta}(\mathbf{q},0)n^{\beta}(-\mathbf{q}),
\end{equation}
where $J_{x,y}=J_\parallel$, $J_z=J_\perp$, and $\chi$ is the spin susceptibility,
\begin{equation}
\chi^{\alpha \beta}(\mathbf{q},i\omega_{l})
=\frac{1}{\beta V}\sum_{i\omega_{n}}\sum_{\mathbf{k}}\text{tr}\left[  G_{0}\left(  \mathbf{k},i\omega_{n}\right)  \sigma^{\alpha}G_{0}\left(  \mathbf{k} +\mathbf{q},i\omega_{n}+i\omega_{l}\right)  \sigma^{\beta}\right],
\end{equation}
and $G_0$ is the Green's function for the Weyl Hamiltonian (\ref{2DHamil}) without the exchange terms.
We obtain
\begin{equation}
H_{\text{DM}}=\int\!d^{2}xD_{\parallel}\left[  n_{z}\text{div}(\mathbf{n}
)-(\mathbf{n}\cdot\mathbf{\nabla})n_{z}\right]  ,\label{DMI}
\end{equation}
with
\begin{equation}
D_{\parallel}=\frac{J_\perp J_\parallel}{8\pi v_\text{F}}\left[  \theta\left(  \left\vert J_\perp \right\vert
+\mu\right)  -\theta\left(  \left\vert J_\perp \right\vert -\mu\right)  \right].\label{eqC}
\end{equation}
The detailed derivation is shown in Supplementary Information \ref{EffecAct}.
It is zero within the band gap of the 2D surface state.
The sign of the DM interaction is positive for $\mu>|J_\perp|$ and negative for $\mu<-|J_\perp|$.
Namely, we can control the sign of the DM interaction by changing the chemical potential by the gate voltage.
It is noted that the sign change of the DM interaction stems from the helicity difference of the momentum--spin locking on the conduction and valence bands.
We evaluate the domain wall energy change due to the DM interaction.
Substituting the domain wall texture (\ref{DW}) into Eq. (\ref{DMI}), we obtain
\begin{equation}
E_\text{DM}=-L D_\parallel \pi \cos \phi,
\end{equation}
with the length of the domain wall $L$.
It takes the minimum energy for the Neel domain wall with $\phi=0$ ($\pi$) for $D_{\parallel}>0$ ($D_{\parallel}<0$).

Finally, we briefly note the general case $0<J_3<J_0$.
When $J_3/J_0$ increases from zero, the exchange interaction on the surface changes from the Ising-like anisotropic form to the Heisenberg-like isotropic form. Therefore, the energy difference among the various domain walls continuously increases.
The numerical results are shown in Supplementary Information \ref{alphadep}.

\textbf{Electron density distribution.}
We demonstrate in Fig. \ref{FigCharge} the electron density distribution of the upper half layers for the minimum-energy domain wall configuration numerically calculated via the expression
\begin{equation}
\rho(x)=\sum_{n\in\text{occupied}} |\psi_n \left(x\right)|^{2},
\end{equation}
where $\left\{\psi_n\right\}$ are the eigenfunctions of the 3D Hamiltonian with the band index $n$.
(For the electron density distributions corresponding to general domain wall configurations, see Supplementary Information \ref{ChargeNB}.)
When $J_3=0$ (Fig. 4(a)), the density distribution is uniform for $\mu=0$, while it is localized at the domain wall for $\mu\neq0$ inside the bulk band gap. The density distribution is inverted between $\mu\longleftrightarrow-\mu$.

\begin{figure}[t]
\centerline{\includegraphics[width=0.3\textwidth]{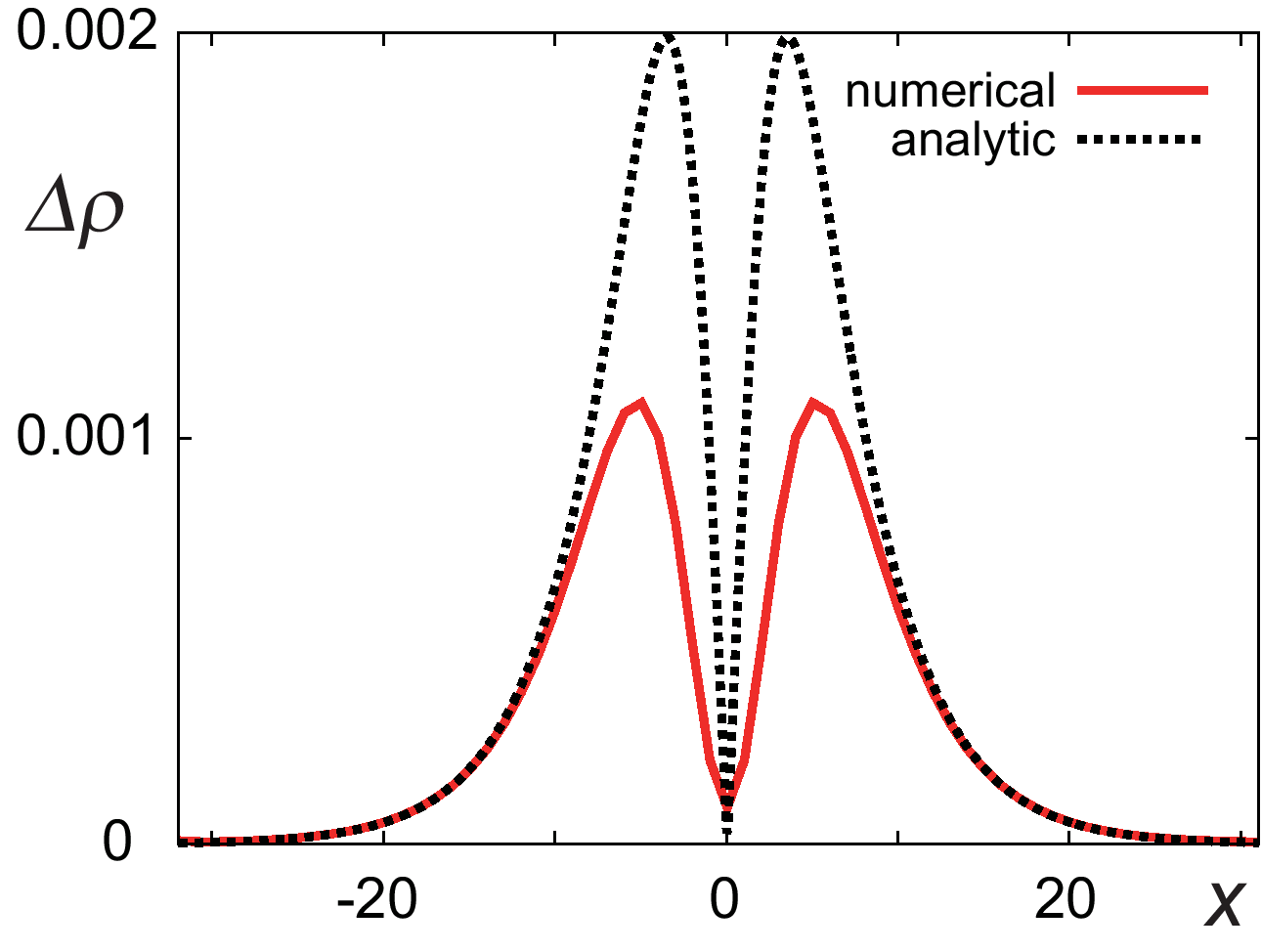}}\caption{Difference of the electron density distribution $\Delta \rho (x)$ (red curve) between the Neel and the Bloch domain wall with equal chemical potential $\mu$.
The horizontal axis is the $x$ coordinate.
It is well explained by the formula Eq. (\ref{ChargeBG}) semi-quantitatively as shown by a black dotted curve. Especially, the dotted curve fits perfectly at tails, where the formula Eq. (\ref{ChargeBG}) is expected to be accurate.}
\label{FigChargeY}
\end{figure}

We may explain the electron accumulation analytically as follows. For $J_3=0$ the
edge state is well described by the Jackiw--Rebbi mode Eq. (\ref{ModeJR}). It gives
the edge channel wave function at zero energy for electrons or holes. When the chemical potential $\mu$ is shifted, the electrons or holes accumulate into the edge
states for $\mu$ within the 2D surface band gap.
Hence, by considering the density of states for the edge channel, the electron density is given by
\begin{equation}
\rho_{\text{JR}}\left(x\right)
=\pm \frac{\mu}{2\pi \hbar v} \left\vert\psi_{\text{JR}}\left(x\right) \right\vert ^2
=\pm C\frac{\mu}{2\pi \hbar v} \left(\cosh\frac{x}{\xi}\right)^{-\left\vert\frac{2\xi}{\hbar v_{\text{F}}}J_{\perp}\right\vert}.
\end{equation}

On the other hand, when $J_3=J_0$, there are two peaks in the density
distribution of the Neel domain wall as found in Fig. \ref{FigCharge}(b).
To understand this behavior, we recall that the electron accumulation due to the spin texture has previously been shown to be~\cite{Nomura2}
\begin{equation}
\rho_{0}\left(  x\right)  =\frac{J_{\parallel}}{2hv_{\text{F}}} \text{sgn}(n_{z})\text{div}[\mathbf{n}\left(  x\right)  ]\label{EqA}
\end{equation}
for a smooth magnetic texture $n_{z}(\neq0)$ which remains almost constant for all over the sample.
Following Ref. \cite{Nomura2}, this relation can be derived by considering the Chern--Simons action
\begin{equation}
S=\frac{1}{2}\sigma_{xy}\int d^3x \varepsilon_{\alpha \beta \gamma}A_\alpha \partial_\beta A_\gamma,
\end{equation}
and the gauge field is
\begin{equation}
A_{x}=\frac{J_{\parallel}}{e\hbar v_{\text{F}}}n_{y},\qquad
A_{y}=-\frac{J_{\parallel}}{e\hbar v_{\text{F}}}n_{x},\qquad
m=J_{\perp}n_{z}.
\end{equation}
By using the relation $\rho_0\left(x\right)=\frac{\delta S}{-e\delta A_0\left(x\right)}$, Eq. (\ref{EqA}) is be obtained.

The total accumulation must consist of the zero-energy
edge contribution $\rho_{\text{JR}}\left(  x\right)  $ and the background
contribution $\rho_{0}\left(  x\right)  $, $\rho(x)=\rho_{\text{JR}}\left(
x\right)  +\rho_{0}\left(  x\right)$.
The Neel-type magnetic configurations contributes to the electron accumulation $\rho_{0}\left(  x\right)$, but that there is no such an accumulation in the Bloch-type magnetic configurations because $n_{x}=0$.
Thus, $\rho_{0}\left(  x\right)$ may be the difference of the electron accumulation between the Neel and Bloch domain walls.
In our case, $\rho_0$ for the Neel wall is
\begin{equation}
\rho_{0}\left(x\right)=\frac{J_{\parallel}}{2hv_{\text{F}}}\frac{\text{sgn}(x)}{\xi}\text{sech}\frac{x}{\xi}\tanh\frac{x}{\xi}\label{ChargeBG}
\end{equation}
with the use of $n_{x}=$sech$(x/\xi), n_z=\tanh(x/\xi)$ in Eq. (\ref{EqA}) for a Neel domain wall.

To confirm this scenario, we plot the difference in the charge density $\Delta \rho (x)$ between the Neel and Bloch walls with the equal chemical potential in Fig. \ref{FigChargeY}. 
The formula Eq. (\ref{ChargeBG}) captures the key structure of the numerical data as in Fig. \ref{FigChargeY}.
Therefore, the peculiar double peak structure in Fig. \ref{FigCharge}(b) stems from the combination of the chiral edge channel and the spatial variation of the spin texture.
The amplitude can be enhanced or reduced, depending on the domain wall type and the filling of the edge channel.

Finally, we note that we can estimate the charging energy with the obtained electron distributions, and conclude that the charging energy is negligible compared with the band energy.
The detailed discussion is shown in Supplementary Information \ref{charging}.

\section*{Discussions}

The origin of the ferromagnetism in doped TI is an important issue.
A first-principles calculation on Mn-doped Bi$_2$Te$_3$~\cite{Henk}
indicates that the Hamiltonian Eq. (\ref{eq:Weyl}) is a good effective model
for the surface states. The gap depends strongly on the direction of 
the magnetization $\bm{M}$; it is $\sim 16$meV when $\bm{M}$ is 
perpendicular to the surface, while the shift in the in-plane momentum 
$\bm{k}_\parallel$ occurs when  $\bm{M}$ is parallel to the surface. 
From the comparison between the gap in the former case 
and the energy shift at $\bm{k}_\parallel =\bm{0}$ in the latter case,
it is concluded that $J_{\parallel} \cong J_\perp $ in 
Eq. (\ref{eq:Weyl}), i.e., $J_3 \cong J_0$. Physically, the Cr and Mn atoms
are replacing Bi, and probably the coupling to the neighboring Te $p$-orbitals
are stronger than to those of Bi atoms, which results in this 
orbital dependent exchange interaction. 
Experimentally, the gating can tune the chemical potential $\mu$
and it has been argued from the dependence of ferromagnetic $T_\text{c}$ on 
$\mu$ that the coupling to the surface Weyl fermions is the origin
of the ferromagnetism~\cite{Joe2}. Therefore, the model Eq. (\ref{eq:Weyl})  
is appropriate also from this viewpoint. 

However, in real materials, the magnetic ions are not selectively doped 
on the surface but are distributed in the whole sample. 
Therefore, it is expected that the magnetization behaves 
uniformly along the $z$ direction (perpendicular to the surface)
for the thin film samples with the thickness of the order of 
8 nm~\cite{Joe}.  The bulk mechanism of ferromagnetism in doped 
TI is studied theoretically also~\cite{Yu,Kurebayashi}.
In Supplementary Information \ref{Layerdep}, we study the dependence on the depth of the magnetic layer.
When the magnetization on the top and bottom surfaces are the same,
the energies of $\phi=0$ domain wall (Neel 1) and $\phi=\pi$ domain wall
(Neel 2) are degenerate because of the mirror symmetry with respect to the plane separating the upper and lower halves of the film. This argument, however, assumes the equivalence 
between the top and bottom surfaces, which is not satisfied in general 
experimental setups. Actually, it is observed that the Weyl points on 
top and bottom surfaces are different in energy typically of the order
of 50meV~\cite{Yoshimi}, and this symmetry is broken.
Therefore, we expect that the type of the domain wall can be 
manipulated by gating.

As for the existence of the domain walls, they are naturally introduced in the 
hysteresis loop in the magnetic field - magnetization curves. 
Actually the longitudinal resistance $R_{xx}$ is found to have the peak
$\sim 2 \frac{h}{e^2}$ at the ends of the hysteresis loop, which is 
likely due to the chiral edge channel associated with the 
domain wall~\cite{Joe}.
 An interesting possibility is the formation of skyrmions, which corresponds to 
the circular closed loop of a domain wall. It is well known  the charge 
doping into $\nu=1$ quantum Hall ferromagnet results in the formation of 
skyrmions~\cite{QHE}.  It remains an open issue if the skyrmions 
can appear in the quantized anomalous Hall system on 3D TI. 
 
\section*{Methods}
We have used the 3D Hamiltonian Eq. (\ref{3DHamil}) for the numerical calculations.
We assume the periodic boundary condition for the $x$ and $y$ directions, and
the open boundary condition for the $z$ direction.
We put non-uniform magnetic moments for the $x$ direction. Therefore, $k_y$
is a good quantum number.
By summing up eigenenergies and amplitudes of eigenfunctions below a
certain particle number, we obtain the total energy and the electron density distribution.
We set the zero of the energy for that of the Bloch wall, and the zero
of the density for that of the half-filling case.
In Fig. \ref{FigChargeY}(c), we compared the density of a Neel wall and
a Bloch wall with the same chemical potential.
We set $t=1, m_0=-0.8, m_2=0.4, J=0.2, \xi=4$ for the main text.

\section*{Acknowledgements}

The authors are grateful for insightful discussions with M. Kawasaki and Y.
Tokura. R. W. and M. E. are grateful for helpful conversations with R. Takahashi and H. Isobe. R. W. was supported by Grant-in-Aid for JSPS Fellows. This work was supported by Grant-in-Aids for Scientific Research (Nos.~24224009, 25400317, and 15H05853) from the Ministry of Education, Culture, Sports, Science and
Technology (MEXT) of Japan.

\section*{Author Contributions}

R. W. performed the numerical calculations. 
R.W., M. E., and N. N. contribute in analyzing the data and writing the paper.

\section*{Competing Financial Interests}

The authors declare that they have no competing financial interests.

\clearpage

\begin{center}
\large{\textbf{Supplementary Information for \\ ``Domain wall of a ferromagnet on a three-dimensional topological insulator''}}
\end{center}

\subsection{Width dependence of the domain wall energy}

\label{WDep}

\begin{figure}[h]
\centerline{\includegraphics[width=0.7\textwidth]{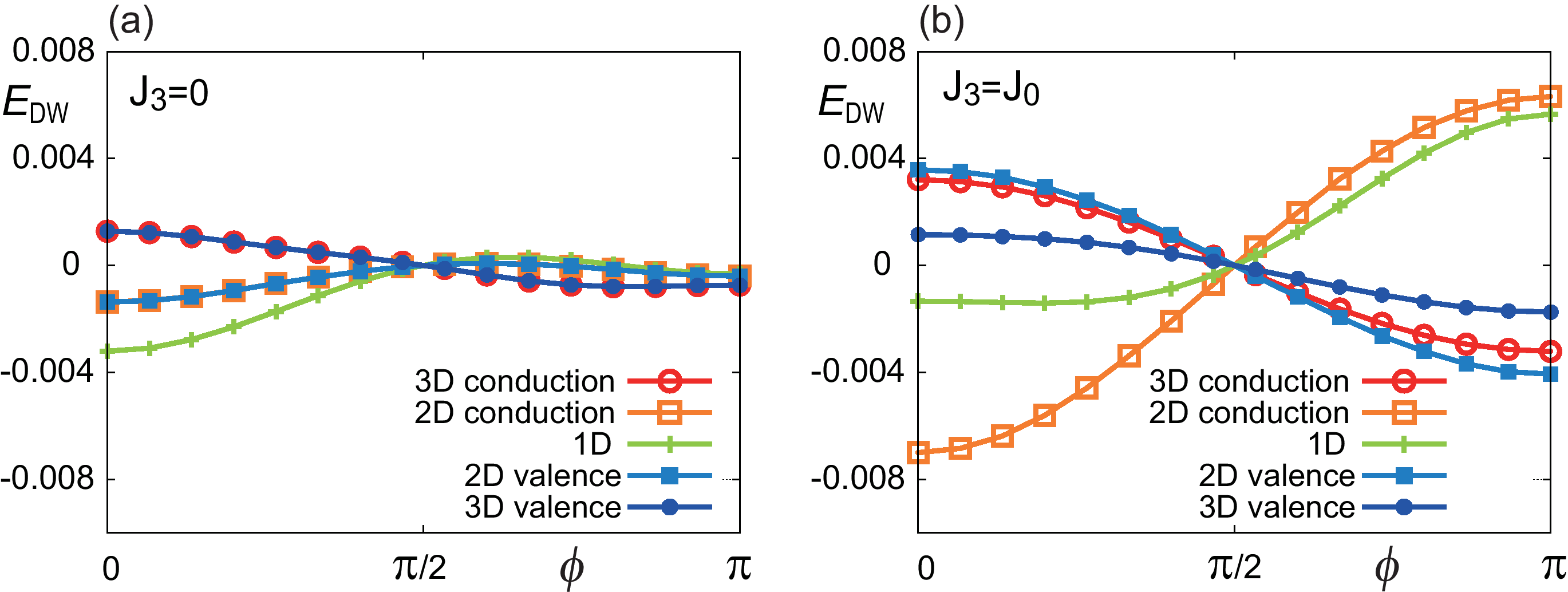}}\caption{Domain
wall energy $E_{\text{DW}}$ as a function of $\phi$ for $\xi=2.0$. The
qualitative behavior does not depend on $\xi$.}
\label{Energyxi2}
\end{figure}

\begin{figure}[h]
\centerline{\includegraphics[width=0.38\textwidth]{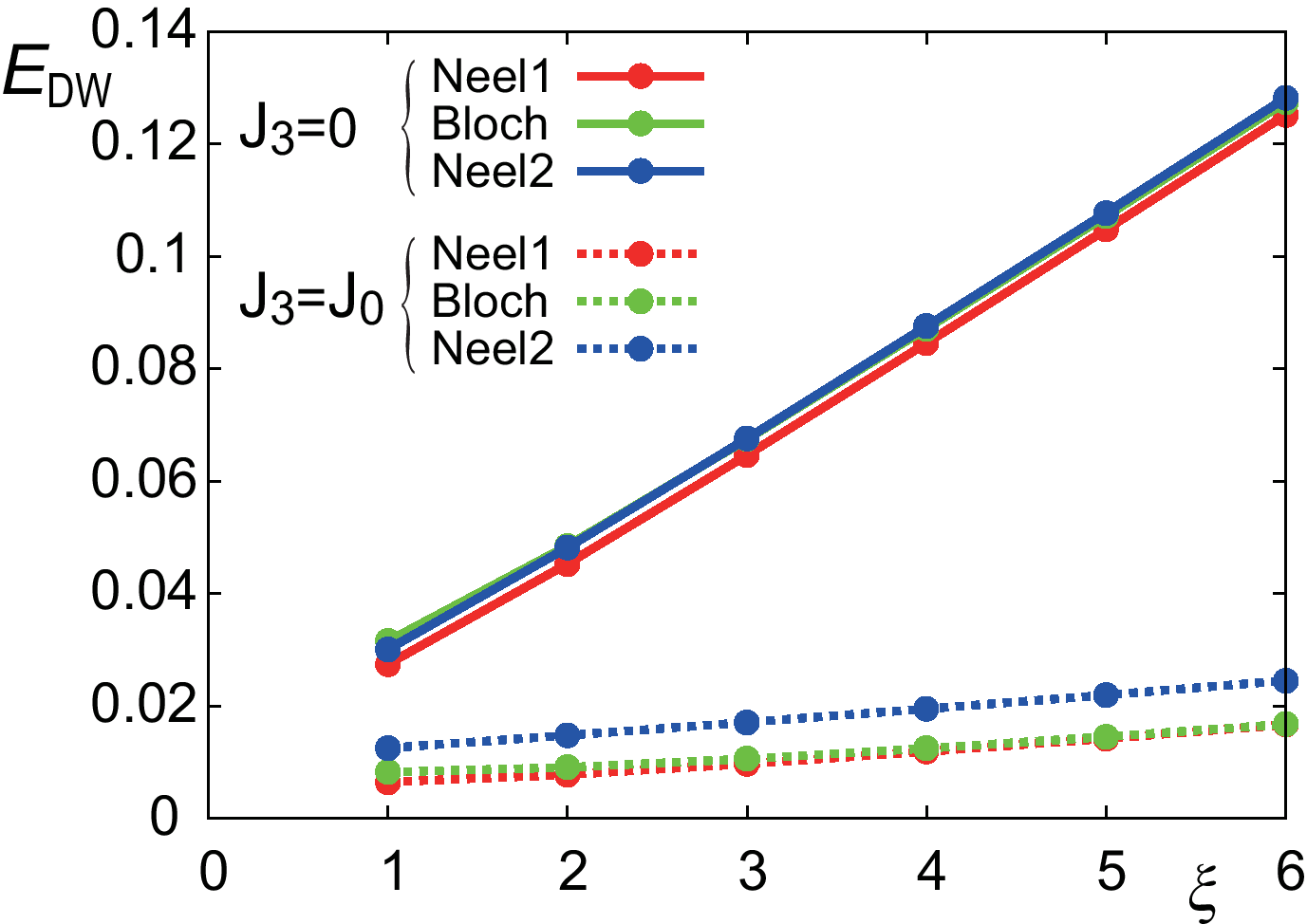}}\caption{Domain
wall energy $E_{\text{DW}}$ as a function of $\xi$ for half-filled case. This
model prefers smaller $\xi$.}
\label{EneXi}
\end{figure}

Figure \ref{Energyxi2} shows the domain wall energy $E_{\text{DW}}$ with the
width $\xi=2.0$ as a function of $\phi$ for $J_{3}=0$ ((a)) and $J_{3}=J_{0}$
((b)). The qualitative behavior is the same as that of Fig. \ref{FigEnergy} for
$\xi=4.0$ in the main text. Figure \ref{EneXi} shows $E_{\text{DW}}$ as a
function of $\xi$. 
It is seen that the smaller $\xi$ is preferred for all the cases. 
For the illustrative purpose, 
we set $\xi=4.0$ in the main text so that we can capture the difference
between various domain wall structures.

\subsection{Effective action}

\label{EffecAct}

We derive an effective action of the 2D Weyl Hamiltonian. The effective
action is obtained by calculating the susceptibility in the one-loop approximation,
\begin{equation}
S_{\text{eff}}=\frac{1}{2}\sum_{\alpha,\beta=x,y,z} J_\alpha J_\beta \sum_{\mathbf{q}} n^{\alpha}(\mathbf{q})\chi^{\alpha \beta}(\mathbf{q},0)n^{\beta}(-\mathbf{q}),
\end{equation}
where $J_{x,y}=J_\parallel$ and $J_z=J_\perp$.
For the region $qv_{\text{F}}\ll |m|$, 
one can expand $\chi^{\alpha \beta}(\mathbf{q},0)$ with respect to $\mathbf{q}$, while for the region $qv_{\text{F}}\gg |m|$, 
the mass $m$ can be put to be zero for the derivation.

We derive the DM interaction and the exchange interaction by calculating the
non-diagonal element of $\chi^{\alpha \beta}$ in the massive Weyl Hamiltonian.
We estimate the domain wall width by calculating the diagonal element of
$\chi^{\alpha \beta}$ in the massless Weyl Hamiltonian.
\newline

\noindent\textbf{The DM and exchange interaction:}

Firstly, we derive the DM interaction and the exchange interaction from the
massive Weyl Hamiltonian. The continuum theory of the 2D Hamiltonian is
described by
\begin{equation}
H=v_{\text{F}}\left( \mathbf{e}_z \times \mathbf{k} \right)\cdot \mathbf{\sigma}+m\sigma_{z}.
\end{equation}
The Green's function reads
\begin{equation}
G_{0}\left(  \mathbf{k},i\omega_{n}\right)  =\frac{1}{i\omega_{n}-\mu-H}
=\frac{\left(  i\omega_n -\mu\right)  \mathbb{I}+v_{\text{F}}\left( \mathbf{e}_z \times \mathbf{k} \right)\cdot \mathbf{\sigma}+m\sigma_{z}}{\left(  i\omega
_{n}-\mu\right)  ^{2}-v_{\text{F}}^{2}\mathbf{k}^{2}-m^{2}}.
\end{equation}
With the use of the Green's function, the susceptibility is given by
\begin{equation}
\chi^{\alpha \beta}(\mathbf{q},i\omega_{l})=\frac{1}{\beta V}\sum_{i\omega_{n}} \sum_{\mathbf{k}}\text{tr}\left[ G_{0}\left(  \mathbf{k}, i\omega_{n}\right)  \sigma^{\alpha}G_{0}\left(  \mathbf{k} +\mathbf{q},i\omega_{n}+i\omega_{l}\right)  \sigma^{\beta}\right].\label{sus}
\end{equation}
We put $i\omega_{l}=0$, and take the $T=0$ limit. We can show
\begin{equation}
\chi^{zx}(\mathbf{q},0)=-i\tilde{D}_{\parallel}q_{x},\quad
\chi^{zy}(\mathbf{q},0)=-i\tilde{D}_{\parallel}q_{y},\quad
\chi^{xz}(\mathbf{q},0)=+i\tilde{D}_{\parallel}q_{x},\quad
\chi^{yz}(\mathbf{q},0)=+i\tilde{D}_{\parallel}q_{y}
\end{equation}
with
\begin{equation}
\tilde{D}_{\parallel}=\frac{1}{4\pi v_{\text{F}}}\left[  \theta\left(  \left\vert
m\right\vert +\mu\right)  -\theta\left(  \left\vert m\right\vert -\mu\right)
\right]  .
\end{equation}
The susceptibility is inverted between $\mu\longleftrightarrow-\mu$. The DM
interaction reads
\begin{equation}
H_{\text{DM}}=D_{\parallel}\int dxdy\left[  n_{z}(\frac{\partial n_{x}}{\partial
x}+\frac{\partial n_{y}}{\partial y})-n_{x}\frac{\partial n_{z}}{\partial
x}-n_{y}\frac{\partial n_{z}}{\partial y}\right]  ,\label{SDMI}
\end{equation}
with $D_{\parallel}=\frac{J_\perp J_\parallel}{2}\tilde{D}_{\parallel}$. This is Eq. (\ref{eqC}) in the main text.

We calculate the diagonal terms $\chi^{\alpha \alpha}(\mathbf{q},0)$ up to $O(q^{2})$,
\begin{equation}
\chi^{xx}(\mathbf{q},0)=-\frac{\Lambda}{3\pi^2 v_{\text{F}}}+\tilde{J}q_{x}^{2},\qquad
\chi^{yy}(\mathbf{q},0)=-\frac{\Lambda}{3\pi^2 v_{\text{F}}}+\tilde{J}q_{y}^{2},\qquad
\chi^{zz}(\mathbf{q},0)=-\frac{\Lambda}{\pi^2 v_{\text{F}}}+\tilde{J}\left(  q_{x}^{2}+q_{y}^{2}\right),
\end{equation}
with
\begin{equation}
\tilde{J}=\frac{1}{12\pi |m|} \left[ \theta\left(\mu+|m|\right) - \theta\left(\mu-|m|\right) \right].
\end{equation}
They contain the ultra-violet momentum cut-off $\Lambda$, which is
naturally given by the 3D band or the lattice constant.
The $q$-independent terms yield
\begin{equation}
H_{K}=-\frac{\Lambda}{3\pi^2 v_F}\int dxdy \left[ J_\parallel^2\left(n_x^2+n_y^2\right)+3J_\perp^2 n_z^2 \right],
\end{equation}
which describes the easy-axis anisotropy, while the $q$-dependent terms yield
\begin{equation}
H_{\text{Ex}}=\tilde{J} \int dxdy \left\{ J_\parallel^2\left[ \left(  \frac{\partial n_{x}}{\partial x}\right)^{2}+\left(  \frac{\partial n_{y}}{\partial y}\right)^{2}\right] + J_\perp^2\left[\left(  \frac{\partial n_{z}}{\partial x}\right)^{2}+\left( \frac{\partial n_{z}}{\partial y}\right)^{2} \right] \right\},
\end{equation}
which acts as the exchange interaction.
\newline

\noindent\textbf{The domain wall width:}

Next, we derive the effective action for $\mu=0,m=0$. We make the following decomposition of Eq. (\ref{sus}),
\begin{equation}
\chi^{\alpha \beta}(\mathbf{q},0)=\chi^{\alpha \beta}(\mathbf{0},0)+\left[
\chi^{\alpha \beta}(\mathbf{q},0)-\chi^{\alpha \beta}(\mathbf{0},0)\right]
.\label{EqX}
\end{equation}
We calculate the $I^{\alpha \beta}(\mathbf{q})\equiv\left[  \chi^{\alpha \beta}(\mathbf{q},0)-\chi^{\alpha \beta}(\mathbf{0},0)\right]  $ term,
\begin{equation}
I^{zz}(\mathbf{q})=\frac{2}{v_{\text{F}}}\int_{0}^{1}dx\int\frac{d^{3}k}{\left(  2\pi\right)  ^{2}}
\frac{\left(  1-x\right)  q^{2}}{\left[k^{2}+x\left(  1-x\right)  q^{2}\right]  ^{2}}=\frac{q}{8v_{\text{F}}},
\end{equation}
and so on. The results for Eq. (\ref{EqX}) read up to $O(q^{2})$ as
\begin{align}
\chi^{xx}(\mathbf{q})&=-\frac{\Lambda}{3\pi^{2}v_{\text{F}}}+\frac{q_{x}^{2}}{16v_{\text{F}}q},\quad
\chi^{yy}(\mathbf{q})=-\frac{\Lambda}{3\pi^{2}v_{\text{F}}}+\frac{q_{y}^{2}}{16v_{\text{F}}q},\quad
\chi^{zz}(\mathbf{q})=-\frac{\Lambda}{\pi^{2}v_{\text{F}}}+\frac{q}{8v_{\text{F}}},\\
\chi^{xy}(\mathbf{q})&=\chi^{yx}(\mathbf{q})=\frac{q_{x}q_{y}}{16v_{\text{F}}\ q}, \quad
\chi^{zx}(\mathbf{q})=\chi^{xz}(\mathbf{q})=\chi^{zy}(\mathbf{q})=\chi^{yz}(\mathbf{q})=0.
\end{align}
The domain wall width $\xi$ is estimated as follows. First, assume that $q \sim 1/\xi$ is much smaller than $|m|/v_F$, and use the expansion Eqs. (S15) and (S17). Then, we demand the anisotropy energy, i.e., the difference between $\chi^{zz}(0,0)$ and $\chi^{xx}(0,0)=\chi^{yy}(0,0)$, is equal to the elastic energy. However, the obtained $q$ is much larger than $|m|/v_F$ since $|m| \gg v_F \Lambda$. Therefore, we need to look for $q \sim 1/\xi$ in the region $q \gg |m|/v_F$, i.e., using Eq. (S21). This results in 
\begin{equation}
\xi=\frac{\pi^{2}}{8\Lambda}\simeq a\label{XI},
\end{equation}
which satisfies $v_F/\xi \gg |m|$ and hence gives the self-consistent estimation. 

\subsection{Edge channel along the domain wall}

\label{EdgeChan}

We investigate the edge channel along the domain wall,
\begin{equation}
n_{x}=\cos\phi\text{sech}\frac{x}{\xi},\quad n_{y}=\cos\phi\text{sech}\frac
{x}{\xi},\quad n_{z}\left(  x\right)  =\tanh\frac{x}{\xi}.
\end{equation}
The exact solution is obtained when $J_{\parallel}=0$, which is known as the
Jackiw-Rebbi solution. The continuum theory of the 2D Hamiltonian in Eq. (\ref{eq:Weyl}) is described by
\begin{equation}
H=v_{\text{F}}\left( \mathbf{e}_z \times \mathbf{k} \right)\cdot \mathbf{\sigma}+J_{\perp}n_{z}(x)\sigma_{z}.
\end{equation}
The eigenequation for $k_{y}=0,E=0$ is
\begin{equation}
\begin{pmatrix}
J_{\perp}n_{z} & -v_{F}\partial_{x}\\
v_{F}\partial_{x} & -J_{\perp}n_{z}
\end{pmatrix}
\begin{pmatrix}
\psi_{\uparrow}(x)\\
\psi_{\downarrow}(x)
\end{pmatrix}
=0.
\end{equation}
The solution which is localized around the domain wall is
\begin{equation}
\psi_{\uparrow,\downarrow}(x)=C\left(  \cosh\frac{x}{\xi}\right)  ^{-J_{\perp
}\frac{\xi}{v_{\text{F}}}},
\end{equation}
which is Eq. (\ref{ModeJR}) in the main text, with the normalization constant
\begin{equation}
C^{2}=\frac{\Gamma\left(  \frac{1}{2}+\frac{J_{\perp}\xi}{v_{\text{F}}}\right)}
{\sqrt{\pi}\xi\Gamma\left(  \frac{J_{\perp}\xi}{v_{\text{F}}}\right)
}.\label{NormaJR}
\end{equation}
In the presence of $n_{x}$ and $n_{y}$ terms for small $J_{\parallel}$, we may
estimate the energy as
\begin{equation}
\bra{\psi}\frac{J_{\parallel}}{\cosh\frac{x}{\xi}}\ket{\psi}
=\frac{J_{\parallel}v_{\text{F}}\Gamma\left(  \frac{1}{2}+\frac{J_{\perp}\xi}{v_{\text{F}}}\right)  ^{2}}{J_{\perp}\xi\Gamma\left(\frac{J_{\perp}\xi}{v_{\text{F}}}\right)  ^{2}},\label{ep}
\end{equation}
which is in Eq. (\ref{edge}) in the main text.

\begin{figure}[h]
\centerline{\includegraphics[width=0.4\textwidth]{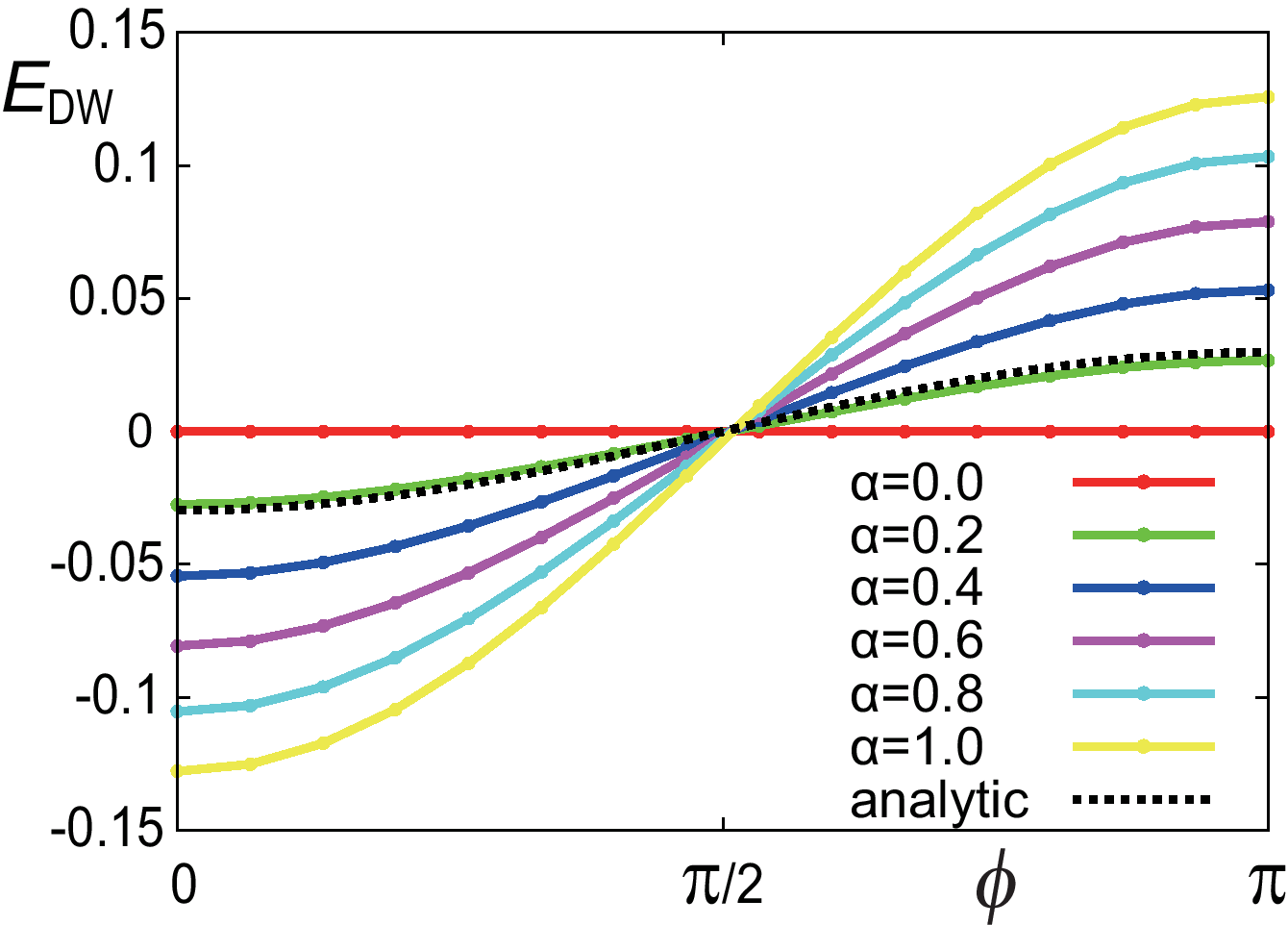}}\caption{Domain
wall energy at $k_{y}=0$ for various $\alpha\equiv J_{3}/J_{0}$. The black
dotted curve is the result of the first-order perturbation for $\alpha=0.2$.}
\label{eneshift}
\end{figure}

Figure \ref{eneshift} shows numerical results on the shift of energy at $k_{y}=0
$ as a function of $\phi$. They are well fitted by the cosine curve obtained
by the first-order perturbation Eq. (\ref{ep}) for $J_{3}/J_{0}=0.2$.

\subsection{$J_{3}$ dependence of the domain wall energy}

\label{alphadep}

\begin{figure}[h]
\centerline{\includegraphics[width=1.0\textwidth]{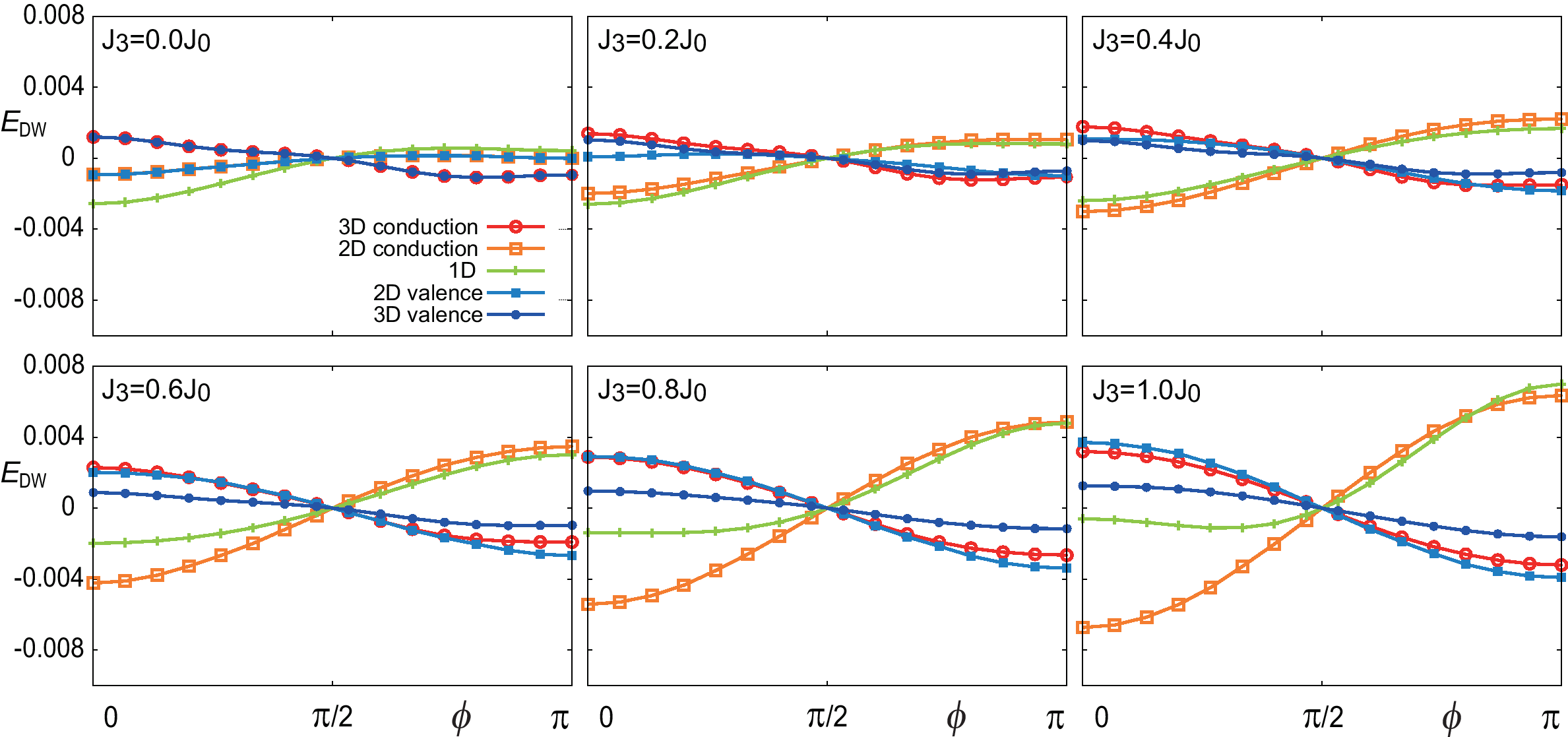}}\caption{Domain
wall energy $E_\text{DW}$ as a function of $\phi$ for various $J_{3}$. }
\label{alpha}
\end{figure}

Figure \ref{alpha} shows the set of figures corresponding to 
Fig. \ref{FigEnergy} in the main text with various $J_{3}$.

\subsection{Symmetry Analysis}

\label{SymmAna}

Here, we summarize the symmetry properties of the models presented in the main text.
\newline

\noindent\textbf{3D Hamiltonian:}

The 3D tight-binding Hamiltonian in Eq. (\ref{3DHamil}) is given by
\begin{equation}
H_{3D}\left(  \mathbf{k},\mathbf{n},J_{0},J_{3},\mu\right)  =t\tau_{x}\left(
\sigma_{x}\sin k_{x}+\sigma_{y}\sin k_{y}+\sigma_{z}\sin k_{z}\right)
+m\tau_{z}+(n_{x}\sigma_{x}+n_{y}\sigma_{y}+n_{z}\sigma_{z})(J_{0}+J_{3}\tau_{z})-\mu,
\end{equation}
where $\mu$ is the chemical potential. Its complex conjugate is
\begin{align}
&  KH_{3D}\left(  \mathbf{k},\mathbf{n},J_{0},J_{3},\mu\right)  K=H_{3D}
^{\ast}\left(  \mathbf{k}\right) \nonumber\\
&  =t\tau_{x}\left(  \sigma_{x}\sin k_{x}-\sigma_{y}\sin k_{y}+\sigma_{z}\sin
k_{z}\right)  +m\tau_{z}+(n_{x}\sigma_{x}-n_{y}\sigma_{y}+n_{z}\sigma
_{z})(J_{0}+J_{3}\tau_{z})-\mu.
\end{align}
We consider an operator
\begin{equation}
\Upsilon=\tau_{x}\sigma_{y},
\end{equation}
which transforms the Hamiltonian as
\begin{equation}
\Upsilon H_{3D}^{\ast}\left(  \mathbf{k},\mathbf{n},J_{0},J_{3},\mu\right)
\Upsilon^{-1}=-H_{3D}\left(  \mathbf{k},\mathbf{n},J_{0},-J_{3},-\mu\right)  .
\end{equation}
On the other hand, the particle-hole operator
\begin{equation}
\Xi=\tau_{y}\sigma_{y}K
\end{equation}
transforms the Hamiltonian as
\begin{equation}
\Xi H_{3D}\left(  \mathbf{k},\mathbf{n},J_{0},J_{3},\mu\right)  \Xi^{-1}
=\tau_{y}\sigma_{y}H_{3D}^{\ast}\left(  \mathbf{k}\right)  \sigma_{y}\tau
_{y}=-H_{3D}\left(  -\mathbf{k},\mathbf{n},J_{0},-J_{3},-\mu\right)
.\label{PHS}
\end{equation}
This operator is the generator of the particle-hole transformation together
with $\mathbf{k}\rightarrow-\mathbf{k}$, $J_{3}\rightarrow-J_{3}$ and $\mu
\rightarrow-\mu$. 
The energy spectrum is symmetric (asymmetric) between the positive and the negative energy when $J_3=0$ ($J_3\neq 0$).

The time-reversal operator is given by
\begin{equation}
\Theta=i\sigma_{y}K,
\end{equation}
which transforms the Hamiltonian as
\begin{equation}
\Theta H_{3D}\left(  \mathbf{k},\mathbf{n},J_{0},J_{3},\mu\right)  \Theta^{-1}
=\sigma_{y}H_{3D}^{\ast}\left(  \mathbf{k}\right)  \sigma_{y}=H_{3D}\left(
-\mathbf{k},-\mathbf{n},J_{0},J_{3},\mu\right)  .
\end{equation}
The chiral symmetry is defined by the product of the particle-hole symmetry
and the time-reversal symmetry,
\begin{equation}
\Pi=\Theta\Xi=i\tau_{y},
\end{equation}
which transforms the Hamiltonian as
\begin{equation}
\Pi H_{3D}\left(  \mathbf{k},\mathbf{n},J_{0},J_{3},\mu\right)  \Pi^{-1}=-H_{3D}\left(
\mathbf{k},-\mathbf{n},J_{0},-J_{3},-\mu\right)  .
\end{equation}
The chiral transformation is preserved 
when $\mathbf{n}=\mathbf{0}$, $J_{0}=0$ and $\mu=0$. 

We make the mirror symmetry along $y$ direction
\begin{equation}
\sigma_{y}\mapsto-\sigma_{y},\quad n_{y}\mapsto-n_{y}.
\end{equation}
The domain wall energy is the same for $\phi$ and $2\pi-\phi$ due to the
mirror symmetry along the $y$ direction as $\sigma_{y}\mapsto-\sigma_{y}$,
$n_{y}\mapsto-n_{y}$. Therefore it is enough to show the results for
$0\leq\phi\leq\pi$.
\newline

\noindent\textbf{2D Hamiltonian:}

The 2D Hamiltonian for the surface state is given by
\begin{equation}
H_{2D}\left(  \mathbf{k},\mathbf{n},\mu\right)  =\hbar v\left(  k_{x}\sigma_{y}
-k_{y}\sigma_{x}\right)  +n_{x}\sigma_{x}+n_{y}\sigma_{y}+n_{z}\sigma_{z}-\mu.
\end{equation}
Its complex conjugate is
\begin{equation}
H_{2D}^{\ast}\left(  \mathbf{k},\mathbf{n},\mu\right)  =\hbar v\left(  -k_{x}\sigma
_{y}-k_{y}\sigma_{x}\right)  +n_{x}\sigma_{x}-n_{y}\sigma_{y}+n_{z}\sigma
_{z}-\mu.
\end{equation}
The time-reversal operator is given by
\begin{equation}
\Theta=i\sigma_{y}K,
\end{equation}
which transforms the Hamiltonian as
\begin{equation}
\Theta H_{2D}\left(  \mathbf{k},\mathbf{n},\mu\right)  \Theta^{-1}=\sigma_{y}
H_{2D}^{\ast}\left(  \mathbf{k}\right)  \sigma_{y}=\hbar v\left(  -k_{x}
\sigma_{y}+k_{y}\sigma_{x}\right)  -n_{x}\sigma_{x}-n_{y}\sigma_{y}
-n_{z}\sigma_{z}-\mu=H_{2D}\left(  -\mathbf{k},-\mathbf{n},\mu\right)  .
\end{equation}
When $\mathbf{n}=\mathbf{0}$, the time-reversal transformation symmetry is preserved,
\begin{equation}
\Theta H_{2D}\left(  \mathbf{k},\mathbf{n}=\mathbf{0},\mu\right)  \Theta^{-1}=H_{2D}\left(
-\mathbf{k},\mathbf{n}=\mathbf{0},\mu\right)  .
\end{equation}
On the other hand, when $\mu=0$, there is a symmetry
\begin{equation}
\Theta H_{2D}\left(  \mathbf{k},\mathbf{n},\mu =0\right)  \Theta^{-1}=-H_{2D}\left(
\mathbf{k},\mathbf{n},\mu =0\right)
\end{equation}
with the time-reversal symmetry operator $\Theta$ even when $\mathbf{n}\neq\mathbf{0}$.

\subsection{Electron accumulation for the Neel wall and the Bloch wall}

\label{ChargeNB}

\begin{figure}[h]
\centerline{\includegraphics[width=0.9\textwidth]{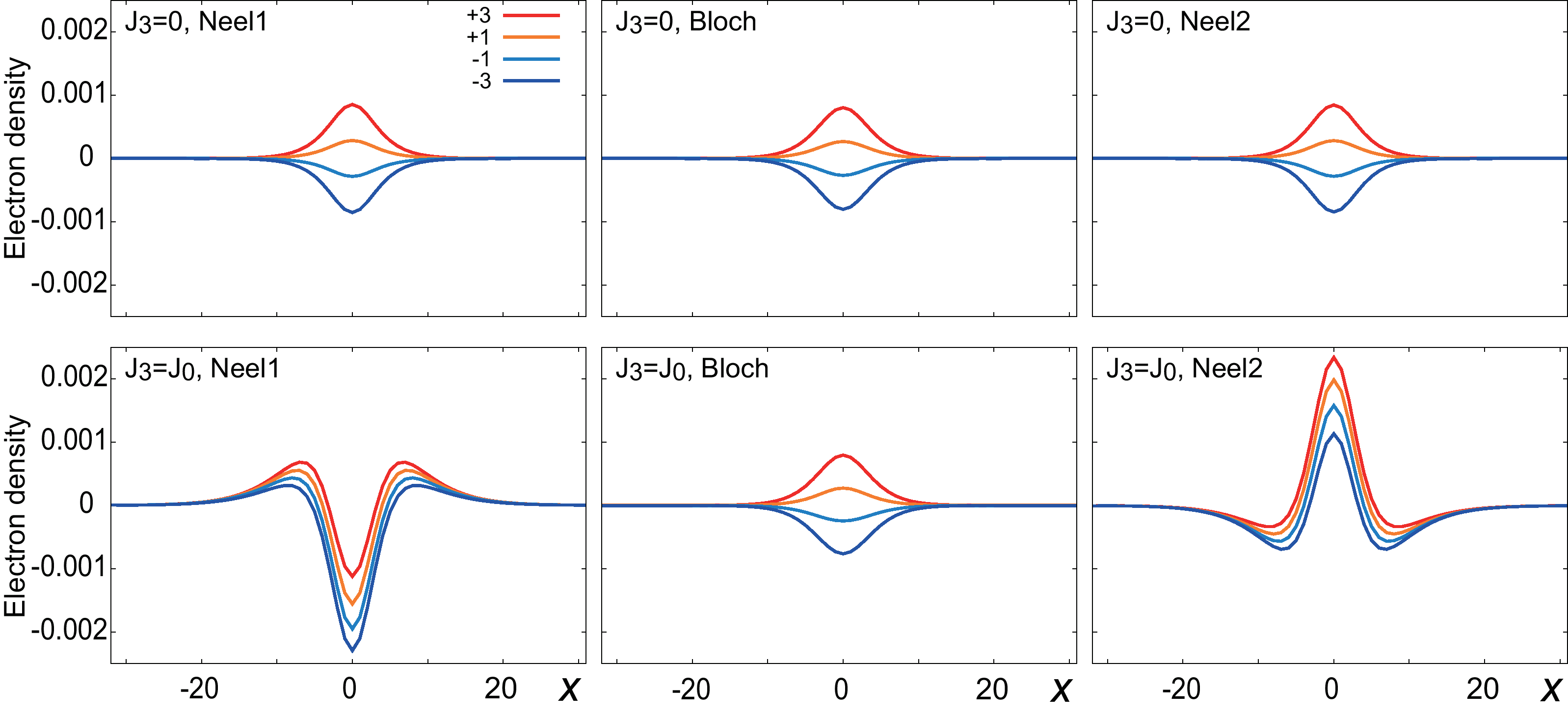}}
\caption{Electron density distribution for various domain wall structures. The electron density distribution is almost the same between the case of $J_{3}=0$ and the Bloch domain wall with $J_{3}=J_{0}$ because the $\mathrm{div}\mathbf{n}$ term is zero. There is a double peak structure for the $J_{3}=J_{0}$ Neel wall due to the $\mathrm{div}\mathbf{n}$ term.}
\label{NBcharge}
\end{figure}

Figure \ref{NBcharge} shows the electron density distribution for the cases of Neel and Bloch walls.
There are two peak structures for $J_{3}=J_{0}$ for the Neel domain wall, which come from $\mathrm{div}\mathbf{n}$ term in Eq. (\ref{EqA}) in the main text. 

\subsection{Charging energy} \label{charging}
We evaluate the charging energy for three cases: (i) heavily doped case where the Fermi energy is in the 2D conduction band, (ii) half-filling case, and (iii) the case where the Fermi energy is deviated from the half-filling but in the 2D bang gap, and show that the charging energy can be neglected compared with the band energy.

We take the lattice constant $a \approx 10$ {\AA} as the unit.
Then, the unit of the Coulomb energy is
\begin{equation}
V = \frac{e^2}{a \varepsilon} = \frac{2 a_B}{\varepsilon a} \frac{e^2}{2a_B} \approx 0.1 \text{eV},
\end{equation}
where, we have used $\varepsilon \approx 20$ for the TIs.

(i) Heavily doped case.
For the heavily doped case, we can neglect the long-range Coulomb interaction and consider only the onsite repulsion because of the screening. According to numerical calculations, the electron distribution amplitudes is less than $10^{-2}$, and the width is about $10$.
Therefore, the charging energy is
\begin{equation}
E_\text{charge} = V \times 2 \times 10 \times \left(10^{-2}\right)^2 = 2 \times 10^{-4} \text{eV},
\end{equation}
where we have taken into account the two domain walls due to the periodic boundary condition.
On the other hand, the order of the band energy is
\begin{equation}
E_\text{band} = 0.004 \times 0.3 = 1.2 \times 10^{-3} \text{eV},
\end{equation}
where we have used $t \approx 3 \text{eV \AA}$ for the TIs.
$E_\text{band}$ is about ten times larger than $E_\text{charge}$. Therefore, we can neglect the effect of the charging energy.

(ii) Half-filling case.
We have to consider the long-range Coulomb interaction.
\begin{align}
E_\text{charge} &= V \int dx dx' \int_0^L dy dy' \frac{\rho(x)\rho(x')}{\sqrt{(x-x')^2+(y-y')^2}} \\
&\approx 2 L V \int dx dx' \left( \log \frac{2L}{|x-x'|} - 1 \right)\rho(x)\rho(x') \\
&\approx 2 L V \rho_0^2 \int_{-\xi/2}^{\xi/2} dx dx' \left( \log\frac{2L}{|x-x'|}-1 \right) \\
&= 2 L V \left( \rho_0 \xi/2 \right)^2 \int_{-1}^1 dx dx' \left( \log \frac{4L}{\xi} - \log |x-x'| -1 \right) \\
&= 2 L V \left( \rho_0 \xi \right)^2 \log \frac{4L}{\xi} + \left( \xi-\text{independent term} \right).
\end{align}

On the other hand, the band energy is calculated as
\begin{equation}
E_\text{band} = \xi L \int_0^{J_\perp} d \varepsilon \varepsilon D \left(\varepsilon\right) = \frac{{J_\perp}^3}{6 \pi v_\text{F}^2} L \xi \equiv \alpha L \xi,
\end{equation}
where we have used the density of states $D \left(\varepsilon\right) = \frac{\varepsilon^2}{2 \pi v_\text{F}^2}$ for the Weyl Hamiltonian.
Therefore, $E_\text{band}$ is proportional to $\xi$ as in the Fig. \ref{EneXi}.
If we use the parameter used in the calculation, we get $\alpha \approx 4 \times 10^{-4}$.
In the Fig. \ref{EneXi}, we obtain $\alpha_1 = 2 \times 10^{-2}$ for the Ising case and $\alpha_2 = 3 \times 10^{-3}$ for the Heisenberg case. In the evaluation, we use the numerical result.

We obtain the optimized $\xi$ by minimizing $E_\text{band} + E_\text{charge}$.
\begin{equation}
\xi_\text{opt} = \frac{2 V \left(\rho_0 \xi\right)^2}{\alpha}.
\end{equation}

For the half-filling case, only the $\nabla \cdot \mathbf{n}$ term contributes the charging, and the amplitude is less than $1/L_y = 1/200$, and $\rho_0 \xi \approx 1/200 \times 10 = 0.05$.
Therefore, we get the optimized $\xi$ as
\begin{equation}
\xi_\text{opt} = \frac{2 \times 2 \times 0.1 \times \left(0.05\right)^2}{\alpha} = 0.3
\end{equation}
for the Heisenberg case.
Therefore, we conclude that the effect of the charging energy is negligible for the half-filling case.

(iii) The case where the Fermi energy is in the 2D gap.
In this case, the Jackiw--Rebbi solutions accumulate.
We can estimate $\rho_0 \xi \approx N / L$ for the $N$ electron doped system.
On the other hand, the particle number is expressed as the function of the Fermi energy.
\begin{equation}
N = \xi L \int_0^\mu d \varepsilon D \left( \varepsilon \right) = \xi L \frac{\mu^2}{4 \pi v_\text{F}^2}.
\end{equation}
Therefore, 
\begin{equation}
\xi_\text{opt} \approx \frac{2 V}{\alpha} \left( \xi^2 \frac{\mu^2}{4 \pi v_\text{F}^2} \right)^2.
\end{equation}
The Fermi energy which gives $\xi_\text{opt} \approx 1$ is $\mu \approx 0.3$eV. It is much larger than the surface gap ($\approx 50$ meV).
Therefore, the charging energy is negligible for the arbitrary Fermi energy.

\subsection{Layer dependence of the domain wall energy}

\label{Layerdep}

\begin{figure}[h]
\centerline{\includegraphics[width=1.0\textwidth]{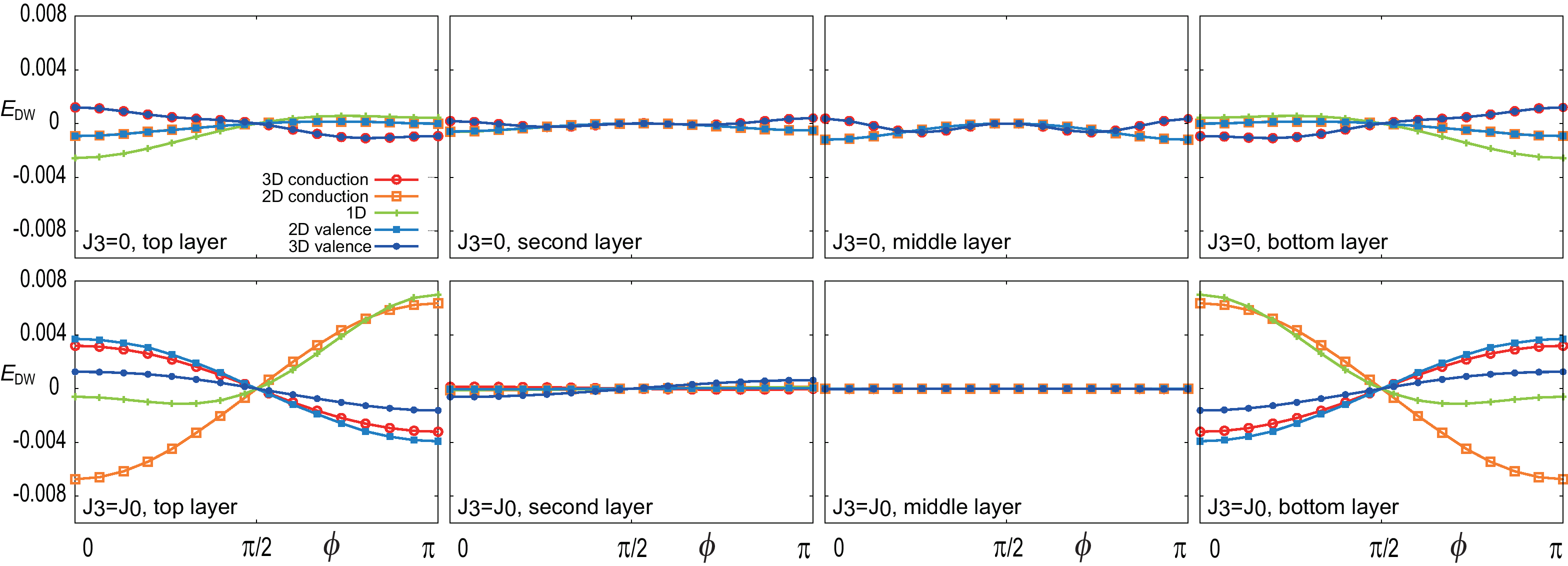}}\caption{Domain wall energy $E_\text{DW}$
for various depths of the magnetic layer.
The figures for the top layer and the
bottom layer are symmetric with respect to $\phi \to \pi - \phi$,
while the figure of the middle layer is symmetric
around $\phi=\pi/2$ because of the mirror symmetry at the middle layer.}
\label{alpha_layer}
\end{figure}

Figure \ref{alpha_layer} shows the domain wall energy $E_\text{DW}$ for various depths of the magnetic layer. 
In the case of
the second and middle layers, the energy differences are small because the
amplitudes of surface states are small. The domain wall energy $E_\text{DW}$
for the top layer with $\phi$ is identical to that of the bottom layer with $\pi
-\phi$, while the domain wall energy of the middle layer is symmetric between
$\phi$ and $\pi-\phi$, because of the mirror symmetry at the middle layer.


\begin{thebibliography}{99}           

\bibitem{TI1}
Hasan, M. Z. \& Kane, C. L.
Colloquium: Topological insulators.
\textit{Rev. Mod. Phys.} \textbf{82}, 3045--3067 (2010).

\bibitem{TI2}
Qi, X.-L. \& Zhang, S.-C.
Topological insulators and superconductors.
\textit{Rev. Mod. Phys.} \textbf{83}, 1057--1110 (2011).

\bibitem{FCZhang}
He, H. T. \textit{et al.}
Impurity effect on weak antilocalization in the topological insulator Bi$_2$Te$_3$.
\textit{Phys. Rev. Lett.} \textbf{106}, 166805 (2011).

\bibitem{Haldane}
Haldane, F. D. M.
Model for a quantum Hall effect without Landau levels: Condensed-matter realization of the ``parity anomaly''.
\textit{Phys. Rev. Lett.} \textbf{61}, 2015--2018 (1988).

\bibitem{Onoda}
Onoda, M \& Nagaosa, N.
Quantized anomalous Hall effect in two-dimensional ferromagnets: Quantum Hall effect in metals.
\textit{Phys. Rev. Lett.} \textbf{90}, 206601 (2003).

\bibitem{Yu}
Yu, R. \textit{et al.}
Quantized anomalous Hall effect in magnetic topological insulators.
\textit{Science} \textbf{329}, 61--64 (2010).

\bibitem{Nomura}
Nomura, K. \& Nagaosa, N.
Surface-quantized anomalous Hall current and the magnetoelectric effect in magnetically disordered topological insulators.
\textit{Phys. Rev. Lett.} \textbf{106}, 166802 (2011).

\bibitem{Abanin}
Abanin, D. A. \& Pesin, D. A.
Ordering of magnetic impurities and tunable electronic properties of topological insulators.
\textit{Phys. Rev. Lett.} \textbf{106}, 136802 (2011).

\bibitem{Franz}
Garate, I. \& Franz, M.
Inverse spin-galvanic effect in the interface between a topological insulator and a ferromagnet.
\textit{Phys. Rev. Lett.} \textbf{104}, 146802 (2010).

\bibitem{Yokoyama}
Yokoyama, T., Zang, J. \& Nagaosa, N.
Theoretical study of the dynamics of magnetization on the topological surface.
\textit{Phys. Rev. B} \textbf{81}, 241410(R) (2010).

\bibitem{Loss}
Tserkovnyak, Y. \& Loss, D.
Thin-film magnetization dynamics on the surface of a topological insulator.
\textit{Phys. Rev. Lett.} \textbf{108}, 187201 (2012).

\bibitem{Linder}
Linder, J.
Improved domain-wall dynamics and magnonic torques using topological insulators.
\textit{Phys. Rev. B} \textbf{90}, 041412(R) (2014).

\bibitem{Ferreiros}
Ferreiros, Y. \& Cortijo, A.
Domain wall motion in junctions of thin-film magnets and topological insulators.
\textit{Phys. Rev. B} \textbf{89}, 024413 (2014).

\bibitem{Baum}
Baum, Y. \& Stern, A.
Density-waves instability and a skyrmion lattice on the surface of strong topological insulators.
\textit{Phys. Rev. B} \textbf{86}, 195116 (2012).

\bibitem{Mendler}
Mendler, B., Kotetes, P. \& Sch\''{o}n, G.
Magnetic order on a topological insulator surface with warping and proximity-induced superconductivity.
\textit{Phys. Rev. B} \textbf{91}, 155405 (2015).

\bibitem{Hurst}
Hurst, H. M., Efimkin, D. K., Zang, J. \& Galitski, V.
Charged skyrmions on the surface of a topological insulator.
\textit{Phys. Rev. B} \textbf{91}, 060401(R) (2015).

\bibitem{Chen}
Chen, Y. L. \textit{et al.}
Massive dirac fermion on the surface of a magnetically doped topological insulator.
\textit{Science} \textbf{329}, 659--662 (2010).
          
\bibitem{Zhang}
Zhang, J. \textit{et al.}
Topology-driven magnetic quantum phase transition in topological insulators.
\textit{Science} \textbf{339}, 1582--1586 (2013).

\bibitem{Chang}
Chang, C.-Z. \textit{et al.}
Experimental observation of the quantum anomalous Hall effect in a magnetic topological insulator.
\textit{Science} \textbf{340}, 167--170 (2013).

\bibitem{Joe}
Checkelsky, J. G. \textit{et al.}
Trajectory of the anomalous Hall effect towards the quantized state in a ferromagnetic topological insulator.
\textit{Nat. Phys.} \textbf{10}, 731--736 (2014).

\bibitem{Chang2}
Chang, C.-Z. \textit{et al.}
High-precision realization of robust quantum anomalous Hall state in a hard ferromagnetic topological insulator.
\textit{Nat. Mater.} \textbf{14}, 473--477 (2015).

\bibitem{Bestwick}
Bestwick, A. J., Fox, E. J., Kou, X., Pan, L. \& Wang, K. L.
Precise quantization of the anomalous Hall effect near zero magnetic field.
\textit{Phys. Rev. Lett.} \textbf{114}, 187201 (2015).

\bibitem{Kou}
Kou, X. \textit{et al.}
Scale-invariant quatnum anomalous Hall effect in magnetic topological insulators beyond the two-dimensional limit.
\textit{Phys. Rev. Lett.} \textbf{113}, 137201 (2014).

\bibitem{Figueroa}
Figueroa, A. I. \textit{et al.}
Magnetic Cr doping of Bi$_2$Se$_3$: Evidence for divalent Cr from x-ray spectroscopy.
\textit{Phys. Rev. B} \textbf{90}, 134402 (2014).

\bibitem{Ni}
Ni, Y., Zhang, Z., Nlebedim, I. C., Hadimani, M. R. \& Tuttle, G. L.
Ferromagnetism of magnetically doped topological insulators in CrxBi2-xTe3 thin films.
\textit{J. Appl. Phys.} \textbf{117}, 17C748 (2015).

\bibitem{Parkin1}
Parkin, S. S. P., Hayashi, M. \& Thomas, L.
Magnetic domain-wall racetrack memory.
\textit{Science} \textbf{320}, 190--194 (2008).

\bibitem{Parkin2}
Ryu, K.-S., Thomas, L., Yang, S.-H. \& Parkin, S. 
Chiral spin torque at magnetic domain walls.
\textit{Nat. Nanotech.} \textbf{8}, 527--533 (2013).

\bibitem{Fert}
Rojas S\'{a}nchez, J. C. \textit{et al.}
Spin-to-charge conversion using Rashba coupling at the interface between non-magnetic materials.
\textit{Nat. Commun.} \textbf{4}, 2944 (2013). 

\bibitem{ZhangH}
Zhang, H., Liu, C.-X., Qi, X.-L., Dai, X., Fang, Z. \& Zhang, S.-C.
Topological insulators in Bi$_2$Se$_3$, Bi$_2$Te$_3$ and Sb$_2$Te$_3$ with a single Dirac cone on the surface.
\textit{Nat. Phys.} \textbf{5}, 438--442 (2009).

\bibitem{Liu}
Liu, C.-X. \textit{et al.}
Model Hamiltonian for topological insulators.
\textit{Phys. Rev. B} \textbf{82}, 045122 (2010).

\bibitem{Shan}
Shan, W.-Y., Lu, H.-Z. \& Shen, S.-Q.
Effective continuous model for surface states and thin films of three-dimensional topological insulators.
\textit{New J. Phys.} \textbf{12}, 043048 (2010).

\bibitem{Henk}
Henk, J. \textit{et al.}
Topological character and magnetism of the Dirac state in Mn-doped Bi$_2$Te$_3$.
\textit{Phys. Rev. Lett.} \textbf{109}, 076801 (2012).

\bibitem{JR}
Jackiw, R. \& Rebbi, C.
Solitons with fermion number 1/2.
\textit{Phys. Rev. D} \textbf{13}, 3398--3409 (1976).

\bibitem{Nomura2}
Nomura, K. \& Nagaosa, N.
Electric charging of magnetic textures on the surface of a topological insulator.
\textit{Phys. Rev. B} \textbf{82}, 161401(R) (2010).

\bibitem{Joe2}
Checkelsky, J. G., Ye, J., Onose, Y., Iwasa, Y. \& Tokura, Y.
Dirac-fermion-mediated ferromagnetism in a topological insulator. 
\textit{Nat. Phys.} \textbf{8}, 729--733 (2012).

\bibitem {Kurebayashi}
Kurebayashi, D. \& Nomura, K.
Weyl semimetal phase in solid-solution narrow-gap semiconductors.
\textit{J. Phys. Soc. Jpn.} \textbf{83}, 063709 (2014).

\bibitem{Yoshimi}
Yoshimi, R. \textit{et al.}
Quantum Hall effect on top and bottom surface states of topological insulator (Bi$_{1-x}$Sb$_x$)$_2$Te$_3$ films.
\textit{Nat. Commun.} \textbf{6}, 6627 (2015).

\bibitem{QHE}
{\it Perspectives in Quantum Hall Effects},
edited by Das Sarma, S. \& Pinczuk, A. 
(Wiley, New York, 1997).


\end{thebibliography}
\end{document}